\documentclass[iop,apj,twocolappendix]{emulateapj-rtx4}
\usepackage{apjfonts}
\usepackage{amssymb,epsfig,mathptmx}

\shortauthors{MAO, KONG, \& LIN}
\shorttitle{FEATURES OF ATTENUATION LAW}

\begin{document}

\title{CHARACTERIZING ULTRAVIOLET AND INFRARED OBSERVATIONAL PROPERTIES FOR GALAXIES. II. FEATURES OF ATTENUATION LAW}

\author{YE-WEI MAO\altaffilmark{1,2,3}, XU KONG\altaffilmark{2,3}, LIN LIN\altaffilmark{4}}

\altaffiltext{1}{Purple Mountain Observatory, Chinese Academy of Sciences, Nanjing 210008, China; ywmao@pmo.ac.cn}
\altaffiltext{2}{Center for Astrophysics, University of Science and Technology of China, Hefei 230026, China; xkong@ustc.edu.cn}
\altaffiltext{3}{Key Laboratory for Research in Galaxies and Cosmology, USTC, CAS, Hefei 230026, China}
\altaffiltext{4}{Shanghai Astronomical Observatory, Chinese Academy of Sciences, Shanghai 200030, China; linlin@shao.ac.cn}

\begin{abstract}
Variations in the attenuation law have a significant impact on observed spectral energy distributions for galaxies. As one important observational property for galaxies at ultraviolet and infrared wavelength bands, the correlation between infrared-to-ultraviolet luminosity ratio and ultraviolet color index (or ultraviolet spectral slope), i.e., the IRX-UV relation (or IRX-$\beta$ relation), offered a widely used formula for correcting dust attenuation in galaxies, but the usability appears to be in doubt now because of considerable dispersion in this relation found by many studies. In this paper, on the basis of spectral synthesis modeling and spatially resolved measurements of four nearby spiral galaxies, we provide an interpretation of the deviation in the IRX-UV relation with variations in the attenuation law. From both theoretical and observational viewpoints, two components in the attenuation curve, the linear background and the 2175 Angstrom bump, are suggested to be the parameters in addition to the stellar population age (addressed in the first paper of this series) in the IRX-UV function; different features in the attenuation curve are diagnosed for the galaxies in our sample. Nevertheless, it is often difficult to ascertain the attenuation law for galaxies in actual observations. Possible reasons for preventing the successful detection of the parameters in the attenuation curve are also discussed in this paper, including the degeneracy of the linear background and the 2175 Angstrom bump in observational channels, the requirement for young and dust-rich systems to study, and the difficulty in accurate estimates of dust attenuations at different wavelength bands.
\end{abstract}


\section{INTRODUCTION}\label{Sec_intro}

The relation between the ratio of infrared (IR) to ultraviolet (UV)
luminosities (i.e., IR excess, or IRX) and the ultraviolet spectral
slope ($\beta$) initially found in starburst galaxies by \citet[][i.e., the
IRX-$\beta$ or IRX-UV relation]{1999ApJ...521...64M} offers a
prescription for correcting internal dust attenuation for galaxies.
This empirical formula has been in particular applied to high
redshift galaxies from which the rest-frame UV emission can be
observed by ground-based optical telescopes
\citep[e.g.,][]{2004ApJ...617..746D, 2007ApJ...670..156D,
2006ApJ...638...72K, 2007ApJS..173..415M, 2009ApJ...705..936B}.
However, the considerable deviation in the IRX-UV relation encountered in studies of normal galaxies on either an integrated or spatially resolved basis casts doubt on the utility of this attenuation diagnostics \citep[e.g.,][]{2002ApJ...577..150B, 2004MNRAS.349..769K, 2005ApJ...619L..51B, 2005ApJ...619L..55S, 2007ApJS..173..392J, 2009ApJ...706..553B}. Differences in the stellar population have been assumed to be a pivotal driver of the deviation in the correlation between IRX and $\beta$ \citep{2004MNRAS.349..769K, 2009ApJ...701.1965M}. However, a number of statistical studies of galaxies adopted relevant spectral indices to trace different stellar populations and did not found any solid trend
of the IRX-UV relation with stellar populations \citep[e.g.,][]{2005ApJ...619L..55S, 2007ApJS..173..392J, 2009ApJ...706..553B}. At the same time, radiative transfer simulations have provided additional insight into the disparities of spectral energy distributions (SEDs) for galaxies with diverse attenuation/extinction properties \citep{2000ApJ...528..799W, 2001PASP..113.1449C, 2006MNRAS.370..380I}. As a result, variations in attenuation/extinction law have been considered to be an additional interpretation of the observed dispersion in the IRX-UV relation \citep{2005MNRAS.360.1413B, 2007MNRAS.375..640P, 2012A&A...539A.145B}.

The terminology "\emph{attenuation}" describes the margin between the intrinsic and observed radiation magnitudes, ascribed to all complex effects of the dust opacity on incident emission, including absorption and scattering by dust grains with the geometric configuration between dust and stars taken into account; as an idealized approximation, homogeneous dust screens in the foreground of stars are often assumed to handle actual observations, and this kind of dust effect is defined as "\emph{extinction}" which correlates only with the optical depth of the dust medium \citep[see][for more thorough discussion about the terminologies]{2001PASP..113.1449C}.

Observations of pointlike sources such as individual stars inside
galaxies are used to study extinction law. Through a spectral
inspection of a set of individual reddened stars from different
sightlines inside our Milky Way (MW), \citet[][hereafter denoted as
FM]{1988ApJ...328..734F, 1990ApJS...72..163F} have obtained a
parameterized description of dust extinction as a function of
wavelength, formulated in terms of five coefficients as follows:
\begin{equation}
\frac{{A(\lambda )}}{{E(\mathrm{B} - \mathrm{V})}} = c_1  + c_2
\lambda ^{ - 1} + c_3 \frac{{\lambda ^{ - 2} }}{{(\lambda ^{ - 2}  -
\lambda _0 ^{ - 2} )^2  + \gamma \lambda ^{ - 2} }} + c_4 F(\lambda
^{ - 1} ) + R_\mathrm{V} , \label{Eq_FM}
\end{equation}
where
\begin{equation}
F(\lambda ^{ - 1} ) = \left\{ \begin{array}{l}
0.5392(\lambda ^{ -
1} - 5.9)^2  + 0.0564(\lambda ^{ - 1}  - 5.9)^3
, \\
~\mathrm{for} ~\lambda ^{ - 1}  \ge 5.9~\mu \mathrm{m}^{ - 1} ,  \\
\\
0, ~\mathrm{for} ~\lambda ^{ - 1}  < 5.9~\mu \mathrm{m}^{ - 1} . \\
\end{array} \right.
\end{equation}
The wavelength $\lambda$ is the free variable in units of $\mu$m.
Equation (\ref{Eq_FM}) characterizes an extinction curve with three
components: a linear background component ($c_1+c_2\lambda^{-1}$), a
bump component ($c_3\lambda^{-2}/((\lambda^{-2}-\lambda
_0^{-2})^2+\gamma\lambda^{-2})$), and a far-UV (FUV) rise component
($c_4F(\lambda^{-1})$). The bump component in the extinction curve was initially detected by \citet{1965ApJ...142.1683S} and is constantly centered at the wavelength of 2175 Angstrom (i.e., $\lambda_0^{-1}=4.596~\mu \mathrm{m}^{-1}$). The mean values of the five coefficients for MW have been found to be $c_1=2.030-3.007c_2$, $c_2=0.698$, $c_3=3.230$, $c_4=0.410$ and $\gamma=0.990$ \citep{1999PASP..111...63F}. A prominent 2175 Angstrom bump is the most appealing property in the MW extinction curve. In Equation (\ref{Eq_FM}), $R_\mathrm{V}(\equiv A(\mathrm{V})/E(\mathrm{B}-\mathrm{V}))$ is the ratio of total to selective extinction, and $R_\mathrm{V}=3.1$ is the best fit for MW. By adjusting the coefficients, the FM parameterization is able to produce a variety of extinction curves with different properties of the three components. At present, the FM formula has been successfully applied not only to our Galaxy but also to our neighbors, including the Magellanic Clouds \citep{2003ApJ...594..279G}. In contrast to our Galaxy, the Small Magellanic Cloud (SMC) presents a lack of a 2175 Angstrom bump and a much steeper linear background in the extinction curve \citep[see][for more details]{2003ApJ...594..279G}.

An alternative form of the extinction curve has been provided by
\citet[][hereafter denoted as CCM]{1989ApJ...345..245C} with a
similar study of individual stars from different sightlines in the MW,
parameterized only with the total-to-selective extinction $R_\mathrm{V}$:
\begin{equation}
\frac{A(\lambda )}{E(\mathrm{B} - \mathrm{V})} = R_\mathrm{V} \cdot
a(\lambda ) + b(\lambda ), \label{Eq_CCM}
\end{equation}
where $a(\lambda)$ and $b(\lambda)$ are piecewise functions of
$\lambda$ in different wavelength ranges \citep[see][for more
details]{1989ApJ...345..245C}. \citet{1994ApJ...422..158O} has updated this $R_\mathrm{V}$-dependent extinction law at optical and near IR wavelength bands. In the UV range ($\lambda^{-1} \geq 3.3~\mu\mathrm{m}^{-1}$, or $\lambda \leq 3030~\mathrm{Angstrom}$), the CCM description is in agreement with the FM parameterization for the MW mean values when $R_\mathrm{V} \sim 3.1$. However, convincing evidence of this $R_\mathrm{V}$-dependent law has not been found for other galaxies, e.g., the Magellanic Clouds \citep{1996ApJ...460..313C, 2003ApJ...594..279G, 2005ApJ...630..355C}, M31 \citep{1992ApJ...400L..35H, 1996ApJ...471..203B}, and M101 \citep{1994A&A...291....1R}.

Different from spatially resolved pointlike sources, in the case of galaxies as a whole, dust mediums are mixed with stars rather than emerging as foreground screens, and the dust obscuration therefore becomes dust attenuation. \citet[][hereafter denoted as C00]{2000ApJ...533..682C} have investigated a sample of local starburst galaxies through integrated measurements and derived an empirical formula of the attenuation curve from polynomial fitting to characterize the complex effects of dust obscuration.
This curve, distinct from those found in the MW and our neighbors, shows a smooth shape and moderate slope without any bump feature. A power-law form of the curve with $\lambda^{-0.7}$ has been suggested to suitably reproduce the C00 polynomial expression for starburst galaxies \citep{2000ApJ...539..718C}. The featureless shape of the C00
attenuation curves is generally believed to be an aftermath of the
statistical approximation and in particular of the "age-selective attenuation" which describes a trend of increasing attenuation for younger stellar populations prevalent in integrated measurements of galaxies \citep{2000ApJ...542..710G, 2007MNRAS.375..640P}. Notwithstanding, the exact form of the attenuation curve for normal galaxies is still unclear. Recent studies have revealed extensive variations in both of the slope and the 2175 Angstrom bump between different galaxies at either low or high redshift \citep[e.g.,][]{2007A&A...472..455N, 2009A&A...499...69N, 2010ApJ...718..184C, 2011ApJ...732..110J, 2011MNRAS.417.1760W, 2011A&A...533A..93B, 2012A&A...545A.141B}.

In the first paper of this series \citep[][hereafter denoted as Paper I]{2012ApJ...757...52M}, we have carried out a spatially resolved study of galaxies, focusing on effects of stellar population age on the IRX-UV relation. In that work, we have divided the measured subregions inside each galaxy into UV clusters representing young stellar populations and diffuse regions in galactic background representing evolved stellar populations. The
age effects appear as systematic offset of the local background regions toward redder colors from the UV clusters in the IRX-UV diagrams; at the same time, complexities in the star formation history (SFH) are suggested to disperse the age effects, and therefore interpret the different levels of overlapping between the UV clusters and the local background regions in the IRX-UV diagrams. However, in Paper I, we assume the C00 attenuation law in the scenarios with dust attenuation and stellar population age addressed as two parameters for interpretation of the observational data. On the basis of this assumption, a few results present a discrepancy between observations and modeling which raises a realistic requirement for more parameters in the IRX-UV function.

As a follow-up of the work in Paper I, in this paper we concentrate on features of the attenuation law in an attempt to find other potential parameters in the IRX-UV function, and we offer a complementary explanation of the observational data. Throughout the paper, consistent with Paper I, IRX is defined as the IR-to-FUV luminosity ratio in the form of a
logarithm, $\mathrm{IRX} \equiv \log(L(\mathrm{IR})/L(\mathrm{FUV}))$, where L(IR) is total IR luminosity radiated by dust in the wavelength range of 3$-$1100 $\mu$m, and $L(\mathrm{FUV})={\nu}L_{\nu}(\mathrm{FUV})$ is the FUV
luminosity of stellar emission. The UV color is defined as
$\mathrm{FUV}-\mathrm{NUV}$, where the FUV and NUV wavebands
correspond to the FUV and NUV channels of the \emph{Galaxy Evolution
Explorer} \citep[\emph{GALEX}; $\lambda_\mathrm{eff}(\mathrm{FUV})=1516~\mathrm{Angstrom}$ and
$\lambda_\mathrm{eff}(\mathrm{NUV})=2267~\mathrm{Angstrom}$;][]{2003lgal.conf...10B,
2005ApJ...619L...1M}.

The reminder of this paper is organized as follows. In Section
\ref{Sec_modeling}, we predict impacts of attenuation curve on the
IRX-UV relation on the basis of modeling. In Section \ref{Sec_results}, we provide an observational insight into the predicted features of the attenuation curve through a spatially resolved study of nearby galaxies. In Section \ref{Sec_mock}, by employing an artificial sample, we assess the reliability of the characterization in previous sections and provide a prescription for quantitative constraints on the attenuation curve. In Section \ref{Sec_disc}, we discuss the main difficulties that potentially prevent a successful detection of the features of the attenuation law in actual observations. Finally, we summarize the results and have a brief outlook on the nature of variations in attenuation law in Section \ref{Sec_sum}.

\section{PREDICTION FROM MODELING}\label{Sec_modeling}

\begin{figure*}[!ht]
\centering
\includegraphics[width=1.8\columnwidth]{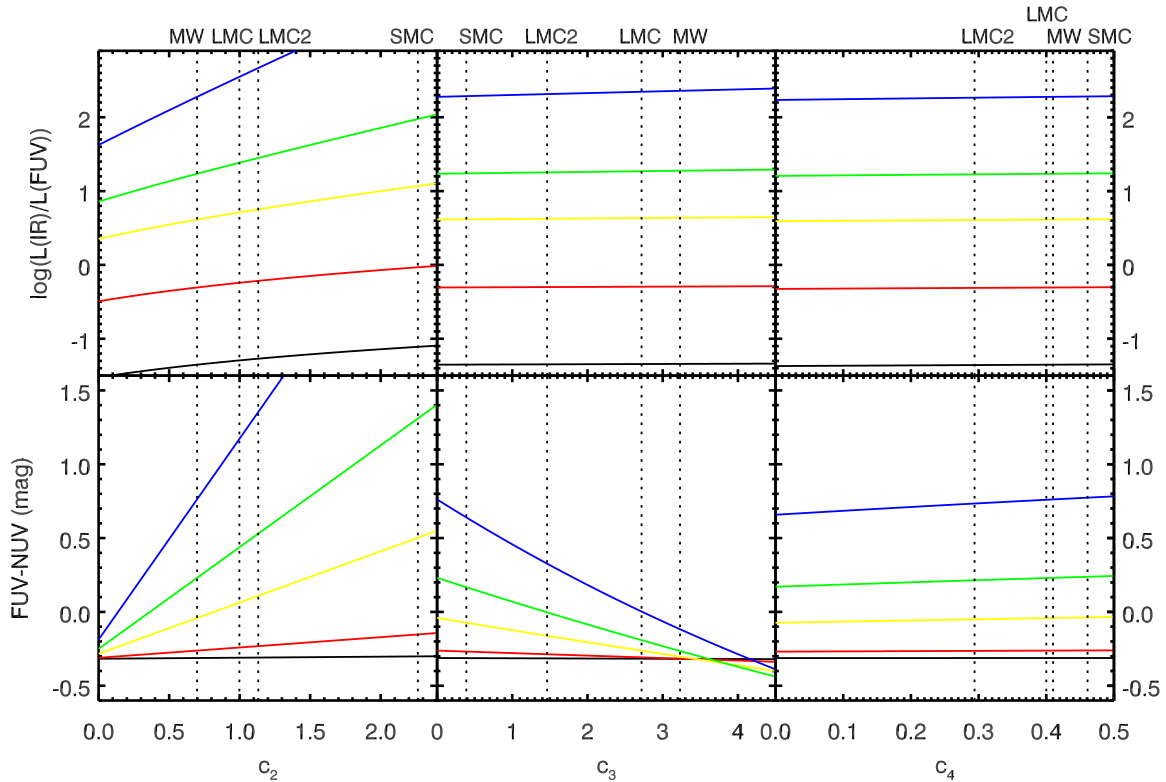}
\caption{IRX (top row) and $\mathrm{FUV}-\mathrm{NUV}$ (bottom row)
vs. $c_2$ (left column), $c_3$ (middle column), and $c_4$ (right column) coefficients in the FM formula. The solid lines in each panel show the modeled correlations at five fixed amounts of dust attenuation: $A$(V) = 0.01 (black), 0.1 (red), 0.5 (yellow), 1.0 (green), and 2.0 (blue) mag; the dotted lines give an example of special fits for the MW, the LMC, the LMC2, and the SMC attenuation curves.}
\label{color_vs_c3}
\end{figure*}

In this section, we carry out a theoretical inspection of possible
diversification in the IRX-UV function related to variations in the
attenuation curve on the basis of modeling with a combination of stellar population synthesis and attenuation curves. We obtain stellar population SEDs with a series of ages from the STARBURST99 spectral synthesis library \citep{1999ApJS..123....3L, 2005ApJ...621..695V}, assuming an instantaneous burst, constant metallicity of $Z=0.02$ and the \citet{2002Sci...295...82K} initial mass function with exponents of 1.3 over 0.1$-$0.5 $M_{\odot}$ and 2.3 over 0.5$-$100 $M_{\odot}$.

Although what we study is attenuation rather than extinction, the FM parameterization (Equation (\ref{Eq_FM})) is employed to construct the attenuation curves in our work. The FM formula is formulated from observations of individual stars in our Galaxy, where the geometric
configuration can be simply described by a homogeneous dust screen in the foreground of stars (attenuation in this case is defined as extinction), whereas in the case of galaxies as a whole, or subregions inside galaxies where individual stars cannot be resolved, actual geometries are more complex than the foreground screen. Nevertheless, the geometrical effects only exert influences on the existing parameters in the attenuation curve such as the linear background slope and the 2175 Angstrom bump strength, and are unlikely to add new parameters. In this situation, the FM formula is valid to reproduce attenuation curves that have been affected by different geometries. We will present a simple test for verifying the validity of the FM parameterization in the case of various dust-star geometries in Appendix \ref{Sec_App1}.

The attenuated spectra in modeling are convolved with transmission curves of the \emph{GALEX} FUV and NUV filters to obtain modeled FUV and NUV magnitudes. The IR luminosity is estimated on the principle of the energy balance between attenuated stellar emission and re-emitted thermal radiation by interstellar dust mediums. In this case, the luminosity difference between the intrinsic and attenuated stellar SEDs can be converted to IR luminosity.

The FM analytical formula parameterizes the three components in the attenuation curve with five coefficients ($c_1$, $c_2$, $c_3$, $c_4$, and $\gamma$, as presented in Equation (\ref{Eq_FM})), which enables a thorough inspection of each component by adjusting the relevant parameters. In this work, we begin with an examination aimed at inspecting the influence of each single component on IRX and $\mathrm{FUV}-\mathrm{NUV}$, respectively. The \emph{GALEX} NUV bandwidth \citep[1771$-$2831 Angstrom;][]{2007ApJS..173..682M} covers the full range of observed widths of the 2175 Angstrom bump \citep[FWHM $\sim$ 360$-$600 Angstrom;][]{1986ApJ...307..286F}, and in this case, the bump area \citep[$A_{\mathrm{bump}} \equiv \pi c_3 /(2 \gamma)$;][]{1986ApJ...307..286F, 2007ApJ...663..320F} becomes the only factor responsible for the impact of the bump on broadband photometric results. Therefore, we fix the width of the 2175 Angstrom bump ($\gamma$) to the MW mean value ($\gamma = 0.99~\mu \mathrm{m}^{-1}$) in this inspection, and then the bump area is solely determined by the coefficient $c_3$. By taking the coefficient $c_1$ in linear correlation with $c_2$ into account \citep{1999PASP..111...63F}, the final variable parameters are $c_2$ (linear background slope), $c_3$ (2175 Angstrom bump strength), and $c_4$ (FUV rise curvature), and we define the values $c_2=0.698$, $c_3=0.0$, and $c_4=0.410$ (i.e., the MW mean values for the linear background slope and the FUV curvature, but no bump) as a standard reference. During the procedure, we vary one coefficient and fix the other two to this reference at each time. This approach allows us to individually inspect the three coefficients, $c_2$, $c_3$, and $c_4$.

\begin{figure*}[!ht]
\centering
\includegraphics[width=1.8\columnwidth]{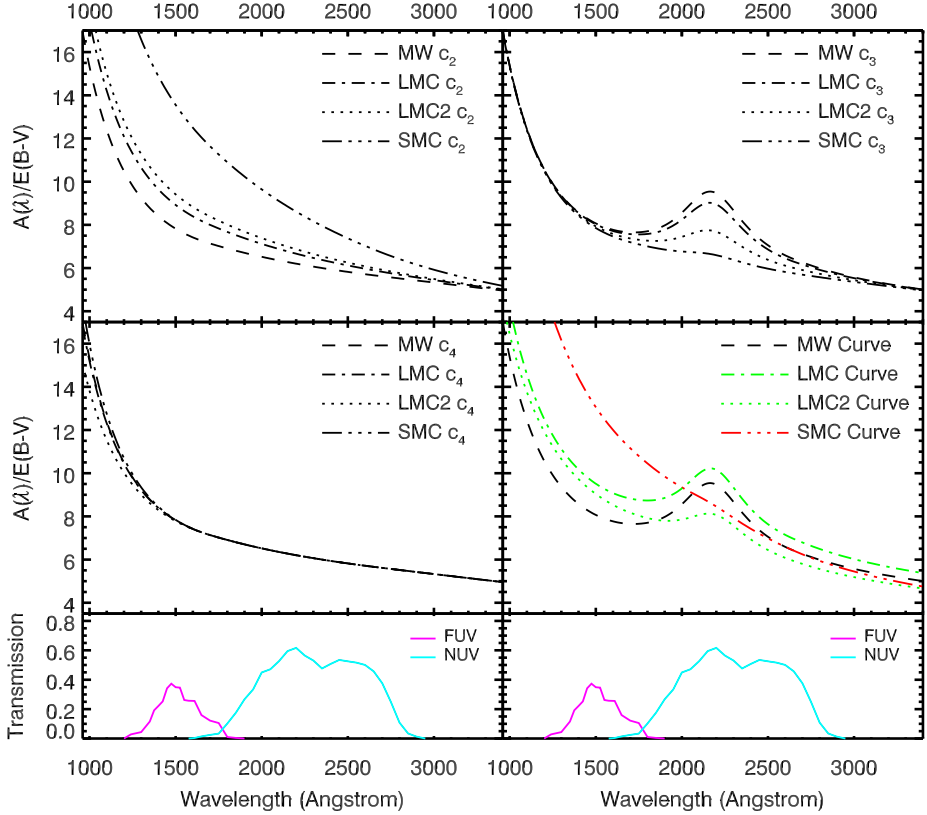}
\caption{Attenuation curves created with Equation (\ref{Eq_FM}) at the
UV wavelength range. The top left panel shows the curves with
different $c_2$: the MW $c_2$ (black dashed line), the LMC $c_2$
(dot-dashed line), the LMC2 $c_2$ (dotted line), and the SMC $c_2$
(triple-dot-dashed line), with $c_3=0.0$ and other coefficients
fixed to the MW values. The top right panel shows the curves with
different $c_3$: the MW $c_3$ (black dashed line), the LMC $c_3$
(dot-dashed line), the LMC2 $c_3$ (dotted line), and the SMC $c_3$
(triple-dot-dashed line), with other coefficients fixed to the MW
values. The middle left panel shows the curves with different $c_4$:
the MW $c_4$ (black dashed line), the LMC $c_4$ (dot-dashed line), the LMC2
$c_4$ (dotted line), and the SMC $c_4$ (triple-dot-dashed line),
with $c_3=0.0$ and other coefficients fixed to the MW values. The
middle right panel shows the MW curve (dashed line), the LMC curve
(dot-dashed line), the LMC2 curve (dotted line), and the SMC curve
(triple-dot-dashed line) curves. Filter transmission curves of the
\emph{GALEX} FUV (magenta) and NUV (cyan) bands are shown in the
bottom panels.}\label{ExtCurve4}
\end{figure*}

The modeling results are plotted in Figure \ref{color_vs_c3} which
shows IRX and $\mathrm{FUV}-\mathrm{NUV}$ as a function of the
coefficients $c_2$, $c_3$, and $c_4$ separately, for five constant
amounts of dust attenuation ($A$(V) = 0.01, 0.1, 0.5, 1.0, and 2.0 mag). Four typical values for the coefficients, i.e., the MW, the SMC, the Large Magellanic Cloud (LMC), and the supershell area of the Large Magellanic Cloud (LMC2), are marked in each panel of Figure \ref{color_vs_c3}, where the MW values are obtained from \citet{1999PASP..111...63F}, and the LMC, LMC2, and SMC values are obtained from \citet{2003ApJ...594..279G}. In this figure, $\mathrm{FUV}-\mathrm{NUV}$ is in positive correlation with $c_2$, and in negative correlation with $c_3$; the amplitude of the variations in $\mathrm{FUV}-\mathrm{NUV}$ tends to increase at heavier attenuation. For instance, in the bottom left panel of Figure \ref{color_vs_c3}, a variation of $c_2$ from 0.698 (the MW value) to 2.264 (the SMC value) reddens $\mathrm{FUV}-\mathrm{NUV}$ by a factor of $\sim 0.6$ mag at $A$(V) = 0.5 mag and $\sim 1.1$ mag at $A$(V) = 1.0 mag; in the bottom middle panel, a variation of $c_3$ from 0.389 (the SMC value) to 3.230 (the MW value) diminishes the reddening in $\mathrm{FUV}-\mathrm{NUV}$ by a factor of $\sim 0.2$ mag at $A$(V) = 0.5 mag and $\sim 0.4$ mag at $A$(V) = 1.0 mag. IRX is in positive correlation with $c_2$, and keeps constant with variations in $c_3$. In contrast to $c_2$ and $c_3$, the coefficient $c_4$ has no effective impact on either IRX or $\mathrm{FUV}-\mathrm{NUV}$.

Figure \ref{ExtCurve4} shows the attenuation curves with $c_2$, $c_3$, and $c_4$ varying at one time, respectively, in the top left, top right, and middle left panels, corresponding to the left, middle, and right columns of Figure \ref{color_vs_c3}. The MW, LMC, LMC2, and SMC values of the three coefficients are employed to actualize the variations. Consistent with the approach for Figure \ref{color_vs_c3}, a single coefficient varies, and the others are set to the standard reference at the same time. In the middle right panel of Figure \ref{ExtCurve4}, the intact MW, LMC, LMC2, and SMC curves are displayed. This exhibition offers more intuitive views of the different properties in the attenuation curve at the FUV and NUV bands. As can be seen, the slope of the linear background preferentially affects the FUV band (1344$-$1786 Angstrom), while the 2175 Angstrom bump dominates in the NUV wavelength range (1771$-$1831 Angstrom). The separate effects of the linear background and the 2175 Angstrom bump on FUV and NUV attenuations in turn clarify the different trends of $\mathrm{FUV}-\mathrm{NUV}$ with $c_2$ and $c_3$ in Figure \ref{color_vs_c3}. This inspection expects the linear background and the 2175 Angstrom bump in the attenuation curve to have conspicuous impacts on IRX and $\mathrm{FUV}-\mathrm{NUV}$, and consequently we will focus on the two components in the following investigations.

\begin{figure*}[!ht]
\centering
\vspace*{-18mm}
\includegraphics[width=1.8\columnwidth]{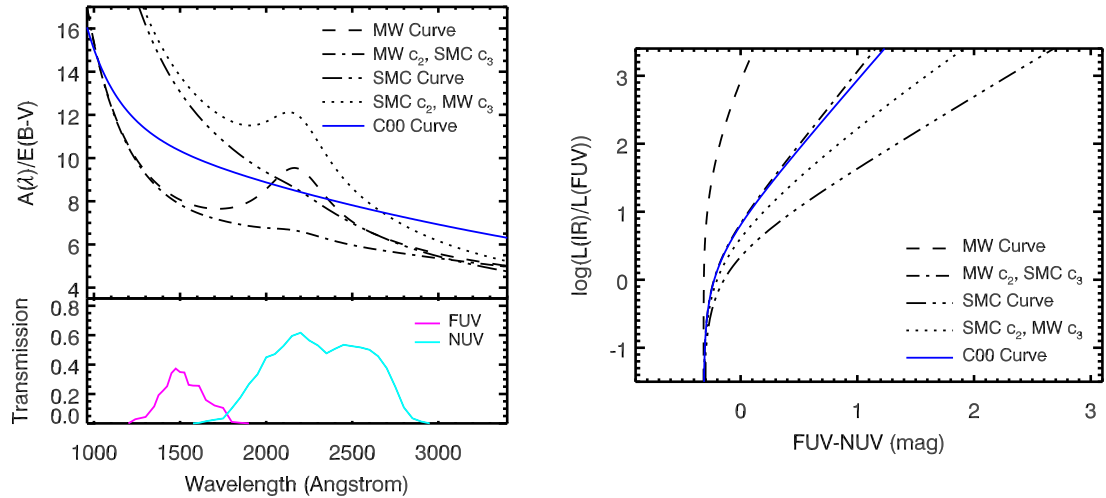}
\vspace*{-23mm}
\caption{Left: attenuation curves with different combinations of
$c_2$ and $c_3$ at UV wavelength range, including the MW curve
(black dashed line), the M-S curve (black dot-dashed line), the SMC
curve (black triple-dot-dashed line), the S-M curve (black dotted
line), and the C00 curves (blue solid line); filter transmission
curves of the \emph{GALEX} FUV (magenta) and NUV (cyan) bands are
also shown. Right: IRX-UV tracks reproduced with the attenuation
curves shown in the left panel.}\label{ExtCurve}
\end{figure*}

In order to better trace the performance of the linear background and the bump in the IRX-UV function, we construct two more attenuation curves: one with the MW-type slope ($c_2=0.698$) and the SMC-type bump ($c_3=0.389$), and the other with the SMC-type slope ($c_2=2.264$) and the MW-type bump ($c_3=3.230$). Combining the two artificial curves with the MW and SMC curves, we have four attenuation curves with different combinations of the two parameters $c_2$ and $c_3$ as a template: the MW curve (representing a shallow slope and a prominent bump),\footnote{Hereafter, we use the "MW curve" to denote the attenuation curve with the MW-type parameters, which could be affected by dust-star geometries and are therefore conceptually different from the MW extinction curve. In the same way, the "SMC curve" described in the text denotes the attenuation curve with the SMC-type parameters.} the curve with the MW $c_2$ and the SMC $c_3$ (hereafter denoted as the M-S curve, representing a shallow slope and a trivial bump), the SMC curve (representing a steep slope and a trivial bump), and the curve with the SMC $c_2$ and the MW $c_3$ (hereafter denoted as the S-M curve, representing a steep slope and a prominent bump). The four attenuation curves are displayed in the left panel of Figure \ref{ExtCurve}, with the C00 curve superimposed for comparison purposes. It is apparent to see that the C00 curve is comparable with the MW curve in slope within the wavelength range of $> 1500 \mathrm{Angstrom}$, albeit with an offset due to the larger value of the total-to-selective extinction for the C00 curve ($R_\mathrm{V} = 4.05$ for the C00 curve, while in the modeling we adopt $R_\mathrm{V} = 3.1$).

The right panel of Figure \ref{ExtCurve} shows the IRX-UV tracks reproduced with the four template attenuation curves and the C00 curve as displayed in the left panel of this figure. In this diagram, all the tracks span a range from $-$1.5 to 3.4 in IRX but extend by various orders of magnitude in $\mathrm{FUV}-\mathrm{NUV}$. For instance, $\mathrm{FUV}-\mathrm{NUV}$ varies by $\sim 0.4$ mag for the MW curve, $\sim 1.4$ mag for the M-S curve, and $\sim 3.0$ mag for the SMC curve. The different extents in $\mathrm{FUV}-\mathrm{NUV}$ on the same IRX scale reflect the distinct features of the linear background and the 2175 Angstrom bump in the IRX-UV function: at constant IRX, the steeper linear background gives rise to redder $\mathrm{FUV}-\mathrm{NUV}$, whereas the presence of the 2175 Angstrom bump tends to restrain the reddening effect. As a consequence, with the bump strength decreasing or the linear background steepening, the IRX-UV relation appears to incline toward the redder color space. Such impacts become of more significance at higher IRX levels (i.e., heavier attenuation). For example, the difference in $\mathrm{FUV}-\mathrm{NUV}$ between the tracks reproduced by the MW curve and the SMC curve ranges from less than 0.2 mag at IRX = 0.0, to about 0.8 mag at IRX = 1.0 and to even over 1.6 mag at IRX = 2.0. It is also worth seeing that in this diagram, the M-S curve and the C00 curve yield consistent IRX-UV tracks, which is ascribed to the similar linear slopes and the same bumpless shape between the M-S curve and the C00 curve (as depicted in the left panel of Figure \ref{ExtCurve}).

It should be noted that in the previous inspection, we vary a single parameter and fix the others at each time. This approach is a simplification in order to illuminate the respective role of each parameter. However, it is suggested that the linear background slope probably depends on size distribution of overall dust grains \citep{2001ApJ...548..296W, 2003ApJ...588..871C}, while the 2175 Angstrom bump could be carried by carbonaceous molecular particles \citep{2003PhRvL..90o5504C}. Although the carbonaceous molecular particles are usually considered to be small grains, there is a lack of conclusive evidence of the link between the grain size distribution and the grain species distribution. Thus, at present we cannot exclude the possibility that the two components in the attenuation curve are likely to change simultaneously and independently in different environments and as a result have composite effects on $\mathrm{FUV}-\mathrm{NUV}$. In the right panel of Figure \ref{ExtCurve}, the IRX-UV track reproduced by the S-M curve is displayed as an example of this degeneracy. Since the presence of the 2175 Angstrom bump decreases the reddening of $\mathrm{FUV}-\mathrm{NUV}$, the track presents bluer $\mathrm{FUV}-\mathrm{NUV}$ than the product of the SMC curve at
constant IRX and therefore cannot be easily distinguished from a shallower attenuation curve with a stronger bump. This degeneracy implies a possible complicacy in the detection of attenuation curve features. In Section \ref{Sec_mock}, we will analyze this degeneracy more thoroughly through an  investigation with an artificial sample.

\section{INSIGHT FROM OBSERVATIONS}\label{Sec_results}

The linear background and the 2175 Angstrom bump in the attenuation curve have been expected to be two important operators in the IRX-UV function in the above section by modeling. In this section, we will examine the IRX-UV relation from an observational standpoint and attempt to find observable phenomena linked to the two parameters in the attenuation curve through a spatially resolved study of four nearby galaxies.

\subsection{Data}\label{Sec_data}

\subsubsection{Sample of Galaxies}

\begin{deluxetable*}{lccccccccc}
\tabletypesize{\tiny} \tablecaption{Basic Properties of the
Galaxies} \tablewidth{0pc} \tablehead{ \colhead{Name} &
\colhead{R.A.\tablenotemark{a}} & \colhead{Dec.\tablenotemark{a}} &
\colhead{Morphology\tablenotemark{a}} &
\colhead{$D_{25}$\tablenotemark{a}} &
\colhead{P.A.\tablenotemark{b}} &
\colhead{Distance\tablenotemark{b}} &
\colhead{$M_{\mathrm{Opt}}$\tablenotemark{a}} &
\colhead{$E$(B-V)$_{\mathrm{GAL}}$\tablenotemark{b,c}} & \colhead{SFR\tablenotemark{a}} \\
\colhead{} & \colhead{(J2000.0)} & \colhead{(J2000.0)} & \colhead{}
& \colhead{(arcmin)} & \colhead{(degree)} & \colhead{(Mpc)} &
\colhead{(mag)} & \colhead{(mag)} & \colhead{($M_{\odot}~\mathrm{yr^{-1}}$)}
 } \startdata
NGC~3031~(M81) & 09~55~33.2 & +69~03~55 & SAab & 26.9~$\times$~14.1 & 157 & 3.7 & $-$21.2 & 0.080 & 1.1 \\
NGC~4536 & 12~34~27.0 & +02~11~17 & SABbc & 7.6~$\times$~3.2 & 130 & 15.0 & $-$20.8 & 0.018 & 3.7 \\
NGC~5194~(M51a) & 13~29~52.7 & +47~11~43 & SABbc & 11.2~$\times$~6.9 & 163 & 7.7 & $-$21.4 & 0.035 & 5.4 \\
NGC~7331 & 22~37~04.1 & +34~24~56 & SAb & 10.5~$\times$~3.7 & 171 & 14.7 & $-$21.8 & 0.091 & 4.2
\enddata
\tablenotetext{a}{\scriptsize ~Data obtained from
\citet{2003PASP..115..928K}.} \tablenotetext{b}{\scriptsize ~Data
obtained from the NASA/IPAC Extragalactic Database.}
\tablenotetext{c}{\scriptsize ~Data obtained from
\citet{1998ApJ...500..525S}.} \label{Tab_sample}
\end{deluxetable*}

In Paper I, we compiled five nearby spiral galaxies from the SINGS \citep{2003PASP..115..928K} sample and discovered influences of stellar population age on the IRX-UV relation. Nevertheless, in that work some observational data lie out of the interpretation with the adopted SFH scenarios that address stellar population age as the second parameter in the IRX-UV function. In this paper, we select four galaxies in Paper I, NGC~3031, NGC~4536, NGC~5194, and NGC~7331, to further investigate the nature of the IRX-UV relation from the alternative viewpoint of attenuation law.\footnote{There are five galaxies studied in Paper I, whereas in this paper NGC~6946 is not selected since the \emph{GALEX} images for this galaxy are quite noisy, and the data points from this galaxy spread with considerable scatter in the IRX-UV diagram as shown in Paper I, which is inappropriate for the current work.} The basic properties of the four galaxies are presented in Table \ref{Tab_sample}.

\subsubsection{Multiwavelength Imaging}

Broadband FUV and NUV imaging data were obtained from \emph{GALEX} observations and downloaded from the Multimission Archive at Space Telescope Science Institute (MAST) Web site;\footnote{{\scriptsize \url{http://galex.stsci.edu/}}} 8 $\mu$m (dust-only \footnote{The 8 $\mu$m data have been transformed into dust-only emission by removing stellar contribution, where the 3.6 $\mu$m data are adopted as the
reference of stellar emission with the scale factor of 0.37 provided in \citet{2007ApJS..173..572T}.}) and 24 $\mu$m images were observed by the \emph{Spitzer Space Telescope} \citep[\emph{Spitzer},][]{2004ApJS..154....1W} and retrieved from the SINGS data distribution service.\footnote{{\scriptsize \url{http://irsa.ipac.caltech.edu/data/SPITZER/SINGS/}}} The original pixel scales are $\sim 1\arcsec.5$ for FUV and NUV, $0\arcsec.75$ for 8 $\mu$m, and $\sim 1\arcsec.5$ for 24 $\mu$m images. The full widths at half maximum (FWHMs) of point spread function (PSF) for these images are $\sim 5\arcsec$ for FUV and NUV, $\sim 2\arcsec$ for 8 $\mu$m, and $\sim 6\arcsec$ for 24 $\mu$m images.

H$\alpha$ narrowband imaging data are also employed in this work. The H$\alpha$ narrowband image for NGC~3031 was observed by the 60/90 cm Schmidt telescope at Xing-Long station of the National Astronomical Observatories of China with the filter of transmission profile FWHM $\sim 120~\mathrm{Angstrom}$ \citep{2003AJ....126.1286L}. The stellar continuum of the H$\alpha$ image was generated through interpolation between two adjacent intermediate-band imaging data centered at 6075 Angstrom and 7050 Angstrom (which were taken with the same telescope for this galaxies) and subtracted from the H$\alpha$ image. The continuum subtraction and flux calibration for the NGC~3031 H$\alpha$ imaging data are presented in \citet{2003AJ....126.1286L}. The H$\alpha$ narrowband images for NGC~4536, NGC~5194, and NGC~7331 were observed by the 2.1 m telescope at Kitt Peak National Observatory with the filter of transmission profile FWHM $\sim 67~\mathrm{Angstrom}$ and downloaded from the SINGS data distribution service. For each of the three galaxies, the stellar continuum of these H$\alpha$ images was produced via rescaling the R band image taken from the same survey, and subtracted from the H$\alpha$ image. Detailed descriptions of the continuum subtraction and flux calibration for these H$\alpha$ images are presented in \citet{2008ApJS..178..247K}. The original pixel scales of the H$\alpha$ images are $\sim 1\arcsec.7$ for NGC~3031, and $\sim 0\arcsec.3$ for NGC~4536, NGC~5194, and NGC~7331. The PSF FWHMs of the H$\alpha$ images are $\sim 4\arcsec.9$ for NGC~3031, and about 1$\arcsec$.4$-$1$\arcsec$.8 for NGC~4536, NGC~5194, and NGC~7331.

The exposure time for all the imaging observations is listed in Table \ref{Tab_exptime}.

\begin{deluxetable}{lrrrrr}
\tablecaption{Total Exposure Time of the Imaging Observations \\(in
units of second)} \tablewidth{0pc} \tablehead{ \colhead{Name} &
\colhead{FUV} & \colhead{NUV} & \colhead{H$\alpha$} & \colhead{8
$\mu$m} & \colhead{24 $\mu$m} } \startdata
NGC~3031 & 14707 & 29422 & 37200 & 6432 & 220 \\
NGC~4536 & 1280 & 1762 & 900 & 1286 & 147 \\
NGC~5194 & 10787 & 10787 & 900 & 2894 & 147 \\
NGC~7331 & 7658 & 12237 & 600 & 1501 & 147
\enddata
\label{Tab_exptime}
\end{deluxetable}

\subsubsection{Image Processing}

The global background is subtracted from the FUV, NUV, and H$\alpha$ images. During this procedure, we mask the galaxy and bright sources in each of the original images and then adopt polynomial fitting along two dimensions in the source-masked images to obtain the global background \citep[see][for the description of global background subtraction in detail]{1999AJ....117.2757Z, 2000AJ....119.2745K, 2011AJ....142...16Z, 2013ApJ...769..127L}. The global background for the 8 $\mu$m and 24 $\mu$m images had already been subtracted before they were released on the Web site.

The PSFs of the FUV, NUV, H$\alpha$, and 8 $\mu$m images are translated to match that of the 24 $\mu$m image via convolution of the images with the relevant kernels offered in \citet{2011PASP..123.1218A}. With regard to the ground-based observations, we apply the "Gauss 5.0 to 24 $\mu$m" kernel to the H$\alpha$ image for NGC~3031 and the "Moffet 1.5 to 24 $\mu$m" kernel to the H$\alpha$ images for NGC~4536, NGC~5194, and NGC~7331 from the same library in the convolution process.

All the PSF-matched images have been registered on the same pixel scale ($\sim 1\arcsec.7$)\footnote{The final pixel scale of $\sim 1\arcsec.7$ is
chosen for the sake of future combination with the ongoing
observations by the 60/90 cm Schmidt telescope at Xing-Long
station of the National Astronomical Observatories of China.} and at
the same coordinate. The image registration is conducted by using the SWarp software \citep{2002ASPC..281..228B}.

\subsubsection{Aperture Photometry}

We select UV clusters representing young stellar populations inside
the galaxies as our targets, which enables us to largely reduce the effects
of stellar population age. The UV clusters are detected as emission
peaks in the FUV images by using the SExtractor software \citep{1996A&AS..117..393B}. Local background subtraction is conducted to remove contamination from diffuse emission associated with evolved stellar populations. We adopt the circular median-filtering method to create local background maps for these galaxies \citep[see][for the description of this approach]{2011AJ....141..205H}. Aperture photometry for the UV clusters is performed in the local-background-subtracted images. Circular apertures are adopted to extract fluxes with the radii of 8.5 arcsec for NGC~3031 and 6.8 arcsec for NGC~4536, NGC~5194, and NGC~7331, corresponding to the physical scales of 152 pc for NGC~3031, 495 pc for NGC~4536, 254 pc for NGC~5194, and 485 pc for NGC~7331, approximately. The aperture size for each galaxy is determined mainly by the spatial resolution of the images: the photometric apertures should be large enough to adequately enclose the resolved objects in each of the images and at the same time small enough to avoid as much emission from other adjacent sources as possible. The UV clusters with the measured fluxes over a 3$\sigma$ level of background deviation in the FUV, NUV, H$\alpha$, 8 $\mu$m, and 24 $\mu$m images are accepted in this work.

According to this selection, more massive clusters are supposed to be more preferentially sampled because of higher luminosities for these objects, and clusters with low mass are possibly undetected because of their luminosities below detection limits; in addition, the selected clusters in more distant galaxies ought to be systematically more massive than those in closer ones, since the photometric apertures enclose larger physical areas for more distant galaxies. As a consequence, the final sample does not contain objects with low mass in the galaxies, particularly for NGC~4536 and NGC~7331 where the mass of the selected clusters is expected to be typically more than $\sim 10^5~M_{\odot}$. In our work, we utilize relative quantities such as luminosity ratio and color index rather than absolute luminosity, and can thus avoid the impact of the mass of the clusters on the results.

As predicted in Section \ref{Sec_modeling}, heavier attenuation allows more effective discovery of the features of the attenuation curve, but the selection of UV clusters has the potential to neglect young but UV-faint regions with large amounts of dust attenuation. In case of missing this kind of sources, we additionally select IR clusters inside the galaxies as emission peaks detected in the 24 $\mu$m images. IR clusters are also good proxies for young stellar populations as a complement of UV clusters especially for dust-obscured regions, since thermal IR radiation in galaxies, particularly with the continuum peaking at around 24 $\mu$m, is triggered by dust heating of star formation activities, and is considered to be a reliable indicator of star formation activities \citep{1998ARA&A..36..189K, 2006ApJ...648..987P, 2007ApJ...666..870C, 2010ApJ...714.1256C}.

Photometry is extracted for the IR clusters in the local-background-subtracted images, and the radii of photometric apertures for each galaxy are the same as in the measurements of the UV clusters. Likewise, we accept the IR clusters with the measured fluxes over a 3$\sigma$ level of background deviation in each of the images. A majority of the IR clusters are the counterparts of the UV clusters with slight displacement in the spatial position,\footnote{A similar displacement between IR and UV emission peaks inside one galaxy has been found in \citet{2005ApJ...633..871C}.} while there are also a few number of independent sources in the IR-selected sample.

Finally, we have 187 UV clusters and 150 IR clusters in NGC~3031, 53 UV clusters and 35 IR clusters in NGC~4536, 100 UV clusters and 117 IR clusters in NGC~5194, and 60 UV clusters and 41 IR clusters in NGC~7331 in our sample. The measured FUV and NUV fluxes for the clusters are corrected for Galactic foreground extinction through the conversion provided in \citet{2007ApJS..173..185G}: $A_{\mathrm{FUV}}=7.9E(\mathrm{B}-\mathrm{V})$ and $A_{\mathrm{NUV}}=8.0E(\mathrm{B}-\mathrm{V})$, with $R_\mathrm{V}=3.1$. The color excess of the Galactic extinction $E(\mathrm{B}-\mathrm{V})$ of each galaxy in our sample is provided by the \citet{1998ApJ...500..525S} Galactic dust map and can be directly obtained from the NASA/IPAC Extragalactic Database.\footnote{{\scriptsize \url{http://ned.ipac.caltech.edu/}}} The measured H$\alpha$ fluxes are corrected for contamination from $[\mathrm{NII}]\lambda\lambda6548,6583$ emission lines by assuming $[\mathrm{NII}]\lambda\lambda6548,6583 / \mathrm{H}\alpha = 0.5$ \citep[as suggested in][for metal-rich galaxies]{2004ApJ...615..228B, 2007ApJ...671..333K}. All the photometric fluxes are corrected for aperture effects, following the instructions of the MIPS instrument handbook provided by the \emph{Spitzer} Science Center.\footnote{{\tiny \url{http://irsa.ipac.caltech.edu/data/SPITZER/docs/mips/mipsinstrumenthandbook/}}} Since we have all the images identically matched with the 24 $\mu$m PSF, the factor for the aperture correction is the same at different bands for each galaxy, depending on the aperture radius. Specifically, the aperture correction factor is 1.50 for NGC~3031, and 1.63 for NGC~4536, NGC~5194, and NGC~7331.

The broadband monochromatic luminosities are calculated
according to the definition $L(\lambda)={\nu}L_{\nu}(\lambda)$. The
narrowband H$\alpha$ luminosity is calculated by using the
conversion $L(\lambda) = c L_{\nu}(\lambda) \Delta \lambda /
\lambda^2$, where $c$ is the speed of light and $\Delta\lambda$ is
the FWHM of transmission profiles of the narrowband filters, according to the instruction in Section 7 of the SINGS user's guide.\footnote{{\scriptsize \url{http://irsa.ipac.caltech.edu/data/SPITZER/SINGS/doc/}}} The total IR luminosity is derived from the 8 $\mu$m and 24 $\mu$m
monochromatic luminosities by using Equation (1) in \citet{2005ApJ...633..871C}.

Uncertainties for the photometry are estimated as a quadratic combination of the deviation of the background and the calibration uncertainties in the relevant images. The deviation of the background is introduced by the process of global and local background subtraction and derived from the global- and
local-background-subtracted images. The calibration uncertainties quoted in the quadrature are 0.05 mag for FUV magnitude and 0.03 mag for NUV magnitude \citep{2007ApJS..173..682M}, 10\% for 8 $\mu$m flux \citep{2007ApJ...655..863D}, and 4\% for 24 $\mu$m flux \citep{2007PASP..119..994E}. The calibration uncertainties of the measured H$\alpha$ fluxes are 8\% for NGC~3031 \citep{2003AJ....126.1286L} and 10\% for NGC~4536, NGC~5194, and NGC~7331 \citep[the SINGS user's guide and][]{2008ApJS..178..247K}.

The measured luminosities from the aperture photometry are listed in
Tables \ref{Tab_UVclu} and \ref{Tab_IRclu} for the UV and IR clusters inside the galaxies respectively.

\begin{deluxetable*}{cccccccc}
\tabletypesize{\scriptsize} \tablecolumns{7}
\tablecaption{Aperture Photometry of the UV Clusters}
\tablewidth{0pc} \tablehead{ \colhead{ID} & \colhead{R.A.\tablenotemark{a}} & \colhead{Dec.\tablenotemark{a}} & \colhead{$L$(FUV)\tablenotemark{b,c,d}} &
\colhead{$L$(NUV)\tablenotemark{b,c,d}} & \colhead{$L$(H$\alpha$)\tablenotemark{b,d}} & \colhead{$L$(8
$\mu$m-dust)\tablenotemark{b,d}} & \colhead{$L$(24
$\mu$m)\tablenotemark{b,d}} \\
\colhead{(Cluster Number)} & \colhead{(J2000.0)} & \colhead{(J2000.0)} & \colhead{(erg~s$^{-1}$)}
& \colhead{(erg~s$^{-1}$)} & \colhead{(erg~s$^{-1}$)} & \colhead{(erg~s$^{-1}$)} &
\colhead{(erg~s$^{-1}$)} \\
} \startdata
NGC~3031-UV001 & 148.853 & 69.219 & 1.19e+40~$\pm$~5.65e+38 & 9.14e+39~$\pm$~2.91e+38 & 4.29e+37~$\pm$~8.08e+36 & 5.71e+38~$\pm$~9.51e+37 & 1.61e+38~$\pm$~2.68e+37 \\
NGC~3031-UV002 & 148.755 & 69.216 & 6.61e+39~$\pm$~3.18e+38 & 4.88e+39~$\pm$~1.95e+38 & 2.48e+38~$\pm$~1.81e+37 & 5.27e+39~$\pm$~5.32e+38 & 2.96e+39~$\pm$~1.21e+38 \\
NGC~3031-UV003 & 148.770 & 69.214 & 1.89e+39~$\pm$~1.11e+38 & 1.27e+39~$\pm$~1.43e+38 & 3.88e+37~$\pm$~7.98e+36 & 1.70e+39~$\pm$~1.86e+38 & 6.07e+38~$\pm$~3.56e+37 \\
\nodata & & & & & & \\
NGC~4536-UV001 & 188.586 & 2.220 & 2.98e+40~$\pm$~1.54e+39 & 2.43e+40~$\pm$~9.39e+38 & 3.79e+37~$\pm$~7.86e+36 & 1.37e+40~$\pm$~2.12e+39 & 2.69e+39~$\pm$~6.73e+38 \\
NGC~4536-UV002 & 188.576 & 2.223 & 4.46e+40~$\pm$~2.20e+39 & 3.72e+40~$\pm$~1.23e+39 & 7.07e+37~$\pm$~9.72e+36 & 1.75e+40~$\pm$~2.39e+39 & 4.88e+39~$\pm$~6.92e+38 \\
NGC~4536-UV003 & 188.579 & 2.221 & 1.05e+41~$\pm$~4.98e+39 & 8.37e+40~$\pm$~2.43e+39 & 2.43e+38~$\pm$~2.43e+37 & 3.54e+40~$\pm$~3.89e+39 & 1.31e+40~$\pm$~8.47e+38 \\
\nodata & & & & & & \\
NGC~5194-UV001 & 202.481 & 47.111 & 6.45e+39~$\pm$~1.08e+39 & 4.97e+39~$\pm$~1.17e+39 & 2.13e+38~$\pm$~3.02e+37 & 1.60e+40~$\pm$~2.93e+39 & 3.89e+39~$\pm$~9.24e+38 \\
NGC~5194-UV002 & 202.516 & 47.265 & 2.33e+40~$\pm$~1.51e+39 & 2.19e+40~$\pm$~1.31e+39 & 2.24e+38~$\pm$~3.08e+37 & 1.71e+40~$\pm$~3.00e+39 & 6.43e+39~$\pm$~9.47e+38 \\
NGC~5194-UV003 & 202.515 & 47.263 & 3.14e+40~$\pm$~1.80e+39 & 2.68e+40~$\pm$~1.38e+39 & 6.60e+38~$\pm$~6.50e+37 & 4.52e+40~$\pm$~5.14e+39 & 2.39e+40~$\pm$~1.32e+39 \\
\nodata & & & & & & \\
NGC~7331-UV001 & 339.257 & 34.481 & 4.08e+40~$\pm$~2.02e+39 & 3.42e+40~$\pm$~1.32e+39 & 8.21e+38~$\pm$~7.14e+37 & 3.26e+40~$\pm$~5.26e+39 & 1.10e+40~$\pm$~1.40e+39 \\
NGC~7331-UV002 & 339.271 & 34.472 & 9.14e+39~$\pm$~7.68e+38 & 8.19e+39~$\pm$~9.36e+38 & 1.66e+38~$\pm$~2.93e+37 & 1.73e+40~$\pm$~4.47e+39 & 4.64e+39~$\pm$~1.34e+39 \\
NGC~7331-UV003 & 339.246 & 34.462 & 1.81e+40~$\pm$~1.07e+39 & 1.38e+40~$\pm$~9.86e+38 & 2.88e+38~$\pm$~3.50e+37 & 3.81e+40~$\pm$~5.62e+39 & 5.07e+39~$\pm$~1.34e+39 \\
\nodata & & & & & &
\enddata
\tablecomments{The aperture radius is 8$\arcsec$.5 for NGC~3031, and 6$\arcsec$.8 for NGC~4536, NGC~5194, and NGC~7331. This table is available in its entirety in the online journal. A portion is shown here for guidance regarding its form and content.}
\tablenotetext{a}{~Position of the apertures on the sky.}
\tablenotetext{b}{~Luminosities measured after local background
subtraction.} \tablenotetext{c}{~Luminosities corrected
for Galactic foreground extinction.} \tablenotetext{d}{~Luminosities corrected for aperture effects.}\label{Tab_UVclu}
\end{deluxetable*}

\begin{deluxetable*}{cccccccc}
\tabletypesize{\scriptsize} \tablecolumns{7}
\tablecaption{Aperture Photometry of the IR Clusters}
\tablewidth{0pc} \tablehead{ \colhead{ID} & \colhead{R.A.\tablenotemark{a}} & \colhead{Dec.\tablenotemark{a}} & \colhead{$L$(FUV)\tablenotemark{b,c,d}} &
\colhead{$L$(NUV)\tablenotemark{b,c,d}} & \colhead{$L$(H$\alpha$)\tablenotemark{b,d}} & \colhead{$L$(8
$\mu$m-dust)\tablenotemark{b,d}} & \colhead{$L$(24
$\mu$m)\tablenotemark{b,d}} \\
\colhead{(Cluster Number)} & \colhead{(J2000.0)} & \colhead{(J2000.0)} & \colhead{(erg~s$^{-1}$)}
& \colhead{(erg~s$^{-1}$)} & \colhead{(erg~s$^{-1}$)} & \colhead{(erg~s$^{-1}$)} &
\colhead{(erg~s$^{-1}$)} \\
} \startdata
NGC~3031-IR001 & 148.749 & 69.237 & 4.31e+39~$\pm$~2.14e+38 & 3.36e+39~$\pm$~1.68e+38 & 4.72e+37~$\pm$~8.18e+36 & 1.31e+39~$\pm$~1.51e+38 & 8.42e+38~$\pm$~4.25e+37 \\
NGC~3031-IR002 & 148.756 & 69.216 & 6.32e+39~$\pm$~3.05e+38 & 4.63e+39~$\pm$~1.90e+38 & 2.39e+38~$\pm$~1.76e+37 & 5.23e+39~$\pm$~5.28e+38 & 2.97e+39~$\pm$~1.21e+38 \\
NGC~3031-IR003 & 148.769 & 69.214 & 1.99e+39~$\pm$~1.15e+38 & 1.33e+39~$\pm$~1.44e+38 & 4.22e+37~$\pm$~8.06e+36 & 1.83e+39~$\pm$~1.98e+38 & 6.62e+38~$\pm$~3.71e+37 \\
\nodata & & & & & & \\
NGC~4536-IR001 & 188.580 & 2.221 & 9.91e+40~$\pm$~4.71e+39 & 8.09e+40~$\pm$~2.36e+39 & 2.38e+38~$\pm$~2.39e+37 & 3.27e+40~$\pm$~3.65e+39 & 1.26e+40~$\pm$~8.34e+38 \\
NGC~4536-IR002 & 188.576 & 2.215 & 4.80e+40~$\pm$~2.35e+39 & 4.35e+40~$\pm$~1.38e+39 & 1.14e+38~$\pm$~1.30e+37 & 5.68e+40~$\pm$~5.91e+39 & 2.42e+40~$\pm$~1.17e+39 \\
NGC~4536-IR003 & 188.581 & 2.207 & 3.19e+40~$\pm$~1.63e+39 & 3.38e+40~$\pm$~1.15e+39 & 1.36e+38~$\pm$~1.48e+37 & 6.95e+40~$\pm$~7.13e+39 & 2.14e+40~$\pm$~1.08e+39 \\
\nodata & & & & & & \\
NGC~5194-IR001 & 202.482 & 47.112 & 6.24e+39~$\pm$~1.07e+39 & 4.88e+39~$\pm$~1.17e+39 & 2.09e+38~$\pm$~2.99e+37 & 1.67e+40~$\pm$~2.97e+39 & 4.05e+39~$\pm$~9.25e+38 \\
NGC~5194-IR002 & 202.509 & 47.269 & 5.56e+39~$\pm$~1.06e+39 & 3.88e+39~$\pm$~1.16e+39 & 1.18e+38~$\pm$~2.53e+37 & 2.42e+40~$\pm$~3.45e+39 & 7.83e+39~$\pm$~9.63e+38 \\
NGC~5194-IR003 & 202.515 & 47.263 & 3.14e+40~$\pm$~1.80e+39 & 2.72e+40~$\pm$~1.38e+39 & 6.23e+38~$\pm$~6.18e+37 & 4.20e+40~$\pm$~4.87e+39 & 2.18e+40~$\pm$~1.26e+39 \\
\nodata & & & & & & \\
NGC~7331-IR001 & 339.287 & 34.342 & 2.90e+40~$\pm$~1.51e+39 & 2.85e+40~$\pm$~1.21e+39 & 2.04e+38~$\pm$~3.09e+37 & 3.37e+40~$\pm$~5.33e+39 & 7.97e+39~$\pm$~1.37e+39 \\
NGC~7331-IR002 & 339.257 & 34.482 & 3.65e+40~$\pm$~1.83e+39 & 3.05e+40~$\pm$~1.25e+39 & 7.67e+38~$\pm$~6.74e+37 & 4.08e+40~$\pm$~5.80e+39 & 1.30e+40~$\pm$~1.43e+39 \\
NGC~7331-IR003 & 339.271 & 34.472 & 8.38e+39~$\pm$~7.49e+38 & 7.66e+39~$\pm$~9.33e+38 & 1.62e+38~$\pm$~2.92e+37 & 1.69e+40~$\pm$~4.46e+39 & 4.64e+39~$\pm$~1.34e+39 \\
\nodata & & & & & &
\enddata
\tablecomments{The aperture radius is 8$\arcsec$.5 for NGC~3031, and 6$\arcsec$.8 for NGC~4536, NGC~5194, and NGC~7331.
This table is available in its entirety in the online journal. A
portion is shown here for guidance regarding its form and content.}
\tablenotetext{a}{~Position of the apertures on the sky.}
\tablenotetext{b}{~Luminosities measured after local background
subtraction.} \tablenotetext{c}{~Luminosities corrected
for Galactic foreground extinction.} \tablenotetext{d}{~Luminosities corrected for aperture effects.} \label{Tab_IRclu}
\end{deluxetable*}

\subsection{Observed IRX-UV Relation}\label{Sec_IRX-UV}

\begin{figure*}[!ht]
\centering
\includegraphics[width=1.8\columnwidth]{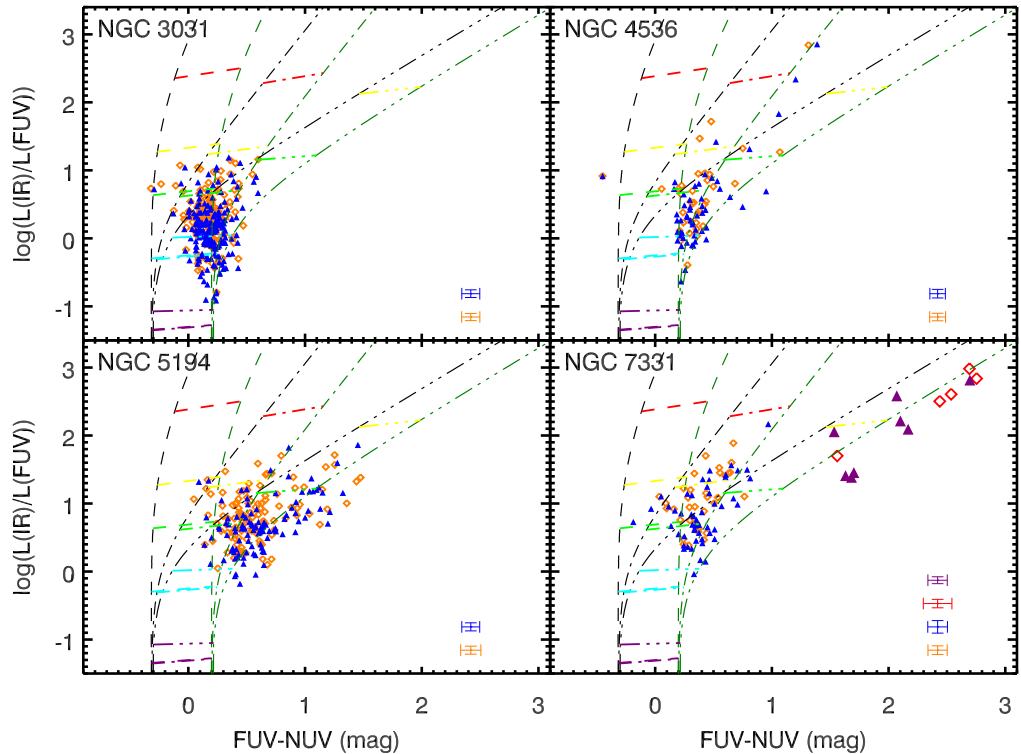}
\caption{IRX vs. $\mathrm{FUV}-\mathrm{NUV}$ for the UV clusters (blue filled triangles) and the IR clusters (orange open diamonds) inside NGC~3031 (top left panel), NGC~4536 (top right panel), NGC~5194 (bottom left panel), and NGC~7331 (bottom right panel). For NGC~7331, the UV and the IR clusters in the central ring are further marked in purple and red, respectively, with larger symbols. The lines superimposed on each panel are obtained by spectral synthesis modeling with different attenuation curves: the MW curve (dashed line), the M-S curve (dot-dashed line), and the SMC curve (triple-dot-dashed line), and they are color-coded by stellar population age: 2 Myr (black) and 100 Myr (dark green). The points of five constant amounts of dust attenuation on the IRX-UV tracks modeled with the same attenuation curve but different stellar population ages are connected by lines with different colors: $A$(V) = 0.01 (purple), 0.1 (cyan), 0.5 (green), 1.0 (yellow), and 2.0 (red) mag. The median errors for the UV clusters (blue) and the IR clusters (orange) are plotted at the bottom right corner of each panel, and for NGC~7331 the error bars are plotted for the disk UV clusters (blue), the disk IR clusters (orange), the ring UV clusters (purple), and the ring IR cluster (red).} \label{IRXUV_gal}
\end{figure*}

With the measurements described above, in this subsection we present the IRX-UV diagrams for each of the galaxies and characterize the results with different signatures of the attenuation curve. Figure \ref{IRXUV_gal} shows the IRX-UV diagrams for the measured clusters inside NGC~3031, NGC~4536, NGC~5194, and NGC~7331, where we adopt the MW, M-S, and SMC curves to offer model comparisons. Two stellar population ages, 2 Myr and 100 Myr, are employed to frame a potential range of age scatter in this relation and describe the envelope of the distribution for objects with the maximal age of 100 Myr in the IRX-UV plane. As shown in this figure, aging from 2 to 100 Myr of simple stellar populations introduces a redward shift by $\sim 0.5$ mag in $\mathrm{FUV}-\mathrm{NUV}$ at fixed IRX.\footnote{In this work, we employ simple stellar populations born with an instantaneous burst in spectral synthesis modeling, in order to illuminate the maximal impact of stellar population age. More complex SFHs are supposed to disperse the age effects, as presented in Paper I.}

In the top left panel of Figure \ref{IRXUV_gal}, the clusters inside NGC~3031 lie in a range from $-$0.2 to 0.5 mag in $\mathrm{FUV}-\mathrm{NUV}$ and from $-$1.0 to 1.0 in IRX, possessing nearly constant $\mathrm{FUV}-\mathrm{NUV}$ space albeit with different IRX. The weak reddening of $\mathrm{FUV}-\mathrm{NUV}$ with increasing IRX coincides with the impact of the 2175 Angstrom bump. The presence of a prominent 2175 Angstrom bump in the attenuation curve provides an interpretation better than the scenarios designed with the bumpless C00 curve in Paper I. However, a majority of the clusters in NGC~3031 are located below IRX $= 0.5$, and the distinction between different attenuation curves is nearly invisible in this range where the deviation in the IRX-UV relation is more likely ascribed to other effects such as the aging of stellar populations.

Contrary to the narrow IRX and color ranges for NGC~3031, the IRX-UV
relation for NGC~4536 presents obvious reddening in $\mathrm{FUV}-\mathrm{NUV}$ with increasing IRX, as shown in the top right panel of Figure \ref{IRXUV_gal}. The clusters span a wide range from 0.0 to 1.4 mag in $\mathrm{FUV}-\mathrm{NUV}$ and from $-$0.7 to 2.9 in IRX. In this diagram, the M-S curve depicts a similar trend to the data distribution. From the right panel of Figure \ref{ExtCurve}, we can see that the M-S curve and the C00 curve reproduce the comparable IRX-UV tracks, and this consistency hints at the reason why the scenarios designed with the C00 curve were applicable to NGC~4536 in Paper I. Furthermore, the tentative analysis based on the three template attenuation curves in this paper implies a bumpless attenuation curve with the MW-type slope for this galaxy.

The reddening inclination of the IRX-UV relation with increasing IRX
can also be seen from the bottom left panel of Figure \ref{IRXUV_gal} for NGC~5194, but the trend appears to be more intensive, i.e., redder $\mathrm{FUV}-\mathrm{NUV}$ at constant IRX than the result from NGC~4536. The data points from NGC~5194 cover the space between the M-S curve and the SMC curve in this diagram. In Paper I, the scenarios with the C00 curve overestimated the stellar population ages for about half of the UV clusters inside NGC~5194, while in this paper, an interpretation with steeper attenuation curves helps to avoid the problem.

\begin{figure}[!ht]
\centering
\hspace*{-21mm}
\includegraphics[width=1.4\columnwidth]{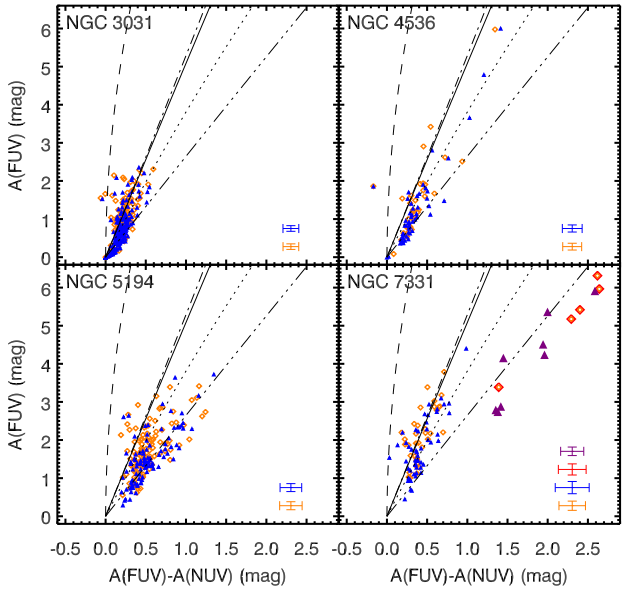}
\caption{FUV attenuation as a function of UV color excess for the
individual galaxies in our sample. The symbols are the same as those
assigned in Figure \ref{IRXUV_gal}. The FUV and NUV attenuations are
estimated on the basis of the B05 calibration. The lines superimposed
on each diagram describe the relations modeled with the MW curve
(dashed line), the M-S curve (dot-dashed line), the SMC curve
(triple-dot-dashed line), the S-M curve (dotted line), and the C00
curve (solid line). The median errors for the UV clusters (blue) and
the IR clusters (orange) are plotted at the bottom right corner of
each panel, and for NGC~7331 the error bars are plotted for the disk
UV clusters (blue), the disk IR clusters (orange), the ring UV
clusters (purple), and the ring IR cluster (red).} \label{comp_Att}
\end{figure}

The bottom right panel of Figure \ref{IRXUV_gal} shows the IRX-UV diagram for NGC~7331. For this galaxy, because of the presence of the IR-prominent ring residing in its central area \citep[][and Paper I]{2004ApJS..154..204R, 2007ApJS..173..572T}, we further divide the measured clusters into disk clusters and ring clusters in order to distinguish their respective features. In this diagram, the disk clusters lie in the range of about 0.0$-$1.0 mag in $\mathrm{FUV}-\mathrm{NUV}$ and 0.0$-$2.0 in IRX, and the ring clusters are located in the range of 1.5$-$2.8 mag in $\mathrm{FUV}-\mathrm{NUV}$ and 1.3$-$3.0 in IRX. The high levels of IRX and the serious reddening of $\mathrm{FUV}-\mathrm{NUV}$ for the ring clusters are assumed to be an aftermath of dust-rich properties in the ring area. It is obvious to see that the disk clusters and the ring clusters possess separate regimes in the IRX-UV plane: the disk clusters appear to follow the relation reproduced by the M-S curve, whereas the ring clusters distribute along the tracks built by the SMC curve. In Paper I, the scenarios modeled with the C00 curve also overestimates stellar population age for the subregions in the ring area of NGC~7331, while in this paper, a steeper attenuation curve such as the SMC curve offers a more suitable interpretation of the data locus.

\subsection{Diagnosis of the Attenuation Curve}\label{Sec_ExtCurv}

We have studied the IRX-UV distributions coincident with the features of the typical attenuation curves by taking each galaxy as a unit in the above section. However, this investigation does not mean that we intentionally presume one constant attenuation curve for a given galaxy. As a matter of fact, the attenuation curve is indeed likely to vary between subregions within one galaxy. In this subsection, we attempt to diagnose the signatures of attenuation curve for each of the galaxies independently of any presumption. A comparison between attenuations at different wavelength bands is the most feasible approach to probing attenuation curve if spectral observations are not available. As the first step, we compare FUV and NUV attenuations ($A$(FUV) and $A$(NUV)) for the measured clusters inside the galaxies.

\begin{figure}[!ht]
\centering
\hspace*{-21mm}
\includegraphics[width=1.4\columnwidth]{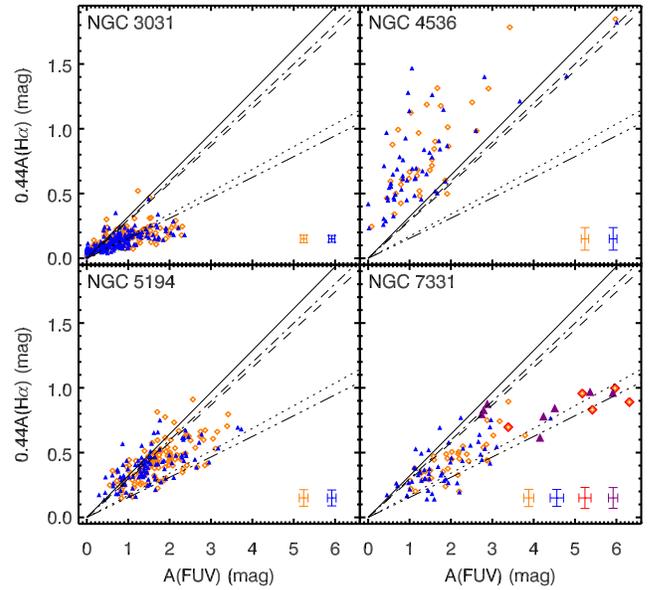}
\caption{H$\alpha$ continuum attenuation as a function of FUV attenuation for the individual galaxies in our sample. The symbols are the same as those assigned in Figure \ref{IRXUV_gal}. The FUV attenuation is estimated on the basis of the B05 calibration, and the H$\alpha$ attenuation is obtained by using the calibration in \citet{2007ApJ...666..870C}. The lines superimposed on each diagram describe the relations modeled with the MW curve (dashed line), the M-S curve (dot-dashed line), the SMC curve (triple-dot-dashed line), the S-M curve (dotted line), and the C00 curve (solid line). The median errors for the UV clusters (blue) and the IR clusters (orange) are plotted at the bottom right corner of each panel, and for NGC~7331 the error bars are plotted for the disk UV clusters (blue), the disk IR clusters (orange), the ring UV clusters (purple), and the ring IR cluster (red).} \label{IRHaUV_Att}
\end{figure}

The observed $A$(FUV) and $A$(NUV) for the clusters are derived from $\log(L(\mathrm{IR})/L(\mathrm{FUV}))$ and $\log(L(\mathrm{IR})/L(\mathrm{NUV}))$, respectively, on the basis of the energy balance principle \citep{1995A&A...293L..65X, 1996A&A...306...61B} and the calibration addressed in \citet[][hereafter denoted as B05]{2005ApJ...619L..51B}:
\begin{equation}
A(\mathrm{FUV}) = - 0.0333 x_1^3 + 0.3522 x_1^2 + 1.1960 x_1 + 0.4967 , \label{Eq_Buat05a}
\end{equation}
and
\begin{equation}
A(\mathrm{NUV}) = - 0.0495 x_2^3 + 0.4718 x_2^2 + 0.8998 x_2 + 0.2269 , \label{Eq_Buat05b}
\end{equation}
where $x_1 \equiv \log(L(\mathrm{IR})/L(\mathrm{FUV}))$ and $x_2
\equiv \log(L(\mathrm{IR})/L(\mathrm{NUV}))$. Equations (\ref{Eq_Buat05a}) and (\ref{Eq_Buat05b}) offer a formula for estimating FUV and NUV attenuations that is not biased by any specific attenuation curve and consequently suitable for the diagnosis of the attenuation curve in our work. The reliability of the B05 formula will be examined in detail in Section \ref{Sec_mock}.

Although there should be no theoretical effect of stellar population age on
the relationship between $A$(FUV) and $A$(NUV), the estimators
$\log(L(\mathrm{IR})/L(\mathrm{FUV}))$ and
$\log(L(\mathrm{IR})/L(\mathrm{NUV}))$ are not definitely
independent of stellar population age \citep{2008MNRAS.386.1157C}. In Figure \ref{IRXUV_gal}, the lines connecting constant attenuation points on the IRX-UV tracks for different ages indicate a very slight difference in IRX between 2 and 100 Myr for the same attenuation; however, the diagnosis based on the comparison between $A$(FUV) and $A$(NUV) can still be possibly affected by this age effect. Via an examination with stellar population synthesis modeling, we find that the application of Equations (\ref{Eq_Buat05a}) and (\ref{Eq_Buat05b}) to stellar populations within an age interval of 2$-$100 Myr would introduce a certain degree of scatter less than $\sim 0.4$ mag in $A(\mathrm{FUV})-A(\mathrm{NUV})$ at fixed $A$(FUV). The influence of this scatter on the diagnosis of the attenuation curve is trivial at a large amount of attenuation but is considerable for objects with low dust content. A more detailed discussion about this effect will be addressed in Section \ref{Sec_disc_age2}.

The comparison between $A$(FUV) and $A$(NUV) is shown in Figure \ref{comp_Att} for the clusters inside NGC~3031, NGC~4536, NGC~5194, and NGC~7331, and the tracks reproduced by the MW curve, the M-S curve, the SMC curve, the S-M curve, and the C00 curve are superimposed on each panel of this figure as well. The modeled lines indicate a trend of decreasing $A$(NUV) at constant $A$(FUV) with weaker bump strength or steeper linear background. A majority of the clusters inside NGC~3031 populate in the range of 0.0$-$2.5 mag in $A$(FUV) and 0.0$-$0.5 mag in $A(\mathrm{FUV})-A(\mathrm{NUV})$, and they appear with quite a small change in $A$(FUV) and $A$(NUV) at the low attenuation level. The clusters inside NGC~4536 present a tight correlation extending to $A(\mathrm{FUV}) \sim 6.0$ mag and $A(\mathrm{FUV})-A(\mathrm{NUV}) \sim 1.5$ mag, and they require a bumpless feature or/and a steeper linear background in the attenuation curve that is different from the MW-type feature to characterize the distribution. Most of the data points from NGC~5194 span about 1.5 mag in $A(\mathrm{FUV})-A(\mathrm{NUV})$ within $A(\mathrm{FUV}) < 4.0$ mag and fall into the transitional space between the M-S curve and the SMC curve, suggesting a bumpless attenuation curve with the slope valuated between the MW and SMC types. For NGC~7331, we can clearly see the different behaviors between the disk and the ring clusters in the diagram: the disk clusters follow the M-S curve in the range of 0.5$-$4.0 mag in $A$(FUV) and 0.0$-$1.0 mag in $A(\mathrm{FUV})-A(\mathrm{NUV})$ approximately, whereas the ring clusters follow the SMC curve at higher levels of both attenuation and reddening, $2.5 < A(\mathrm{FUV}) < 6.3$ mag and $1.2 < A(\mathrm{FUV})-A(\mathrm{NUV}) < 2.7$ mag.

As we have implied at the end of Section \ref{Sec_modeling}, there
is a degeneracy of the linear background and the 2175 Angstrom bump in NUV attenuation, which makes it impossible to estimate the attenuation curve by adopting only UV wavelength bands. The track reproduced by the S-M curve is displayed in Figure \ref{comp_Att} as an example of this degeneracy to illustrate the same appearance produced by different combinations of the linear background slope and the bump strength in the $A$(FUV) versus $A$(NUV) diagnostics. In this situation, it is necessary to have more observational tracers of the parameters in the attenuation curve.

The H$\alpha$ emission line resides at the wavelength of $\lambda~\sim6563~\mathrm{Angstrom}$ beyond the coverage of the 2175 Angstrom bump; therefore, it is in a good position to offer a complementary diagnosis, especially for the linear background slope.

H$\alpha$ emission lines are ionized from gaseous nebulae by young
and massive stars in star-forming regions. The same star formation
activities also heat interstellar dust grains to high temperatures
($>$ 100 K) and excite IR continua peaking at around 24 $\mu$m from the dust. The common association of H$\alpha$ and 24 $\mu$m emission with star formation leads to a correlation between intrinsic (unattenuated) H$\alpha$ and 24 $\mu$m luminosities, i.e., if there is no attenuation, the ratio of 24 $\mu$m to H$\alpha$ luminosities is supposed to be constant. As a result, the combination of observed (attenuated) H$\alpha$ with 24 $\mu$m luminosities is proposed as a robust estimator of H$\alpha$ attenuation, i.e., theoretically, the difference of the observed 24 $\mu$m-to-H$\alpha$ ratio from its intrinsic value is ascribed to H$\alpha$ attenuation \citep{2006ApJ...648..987P, 2007ApJ...666..870C, 2007ApJ...667L.141R, 2007ApJ...671..333K, 2009ApJ...703.1672K}. Compared with UV or optical blue-band attenuation, H$\alpha$ attenuation is insensitive to variations in the attenuation curve since the differences between various attenuation curves are quite slight in long wavelength bands. This property implies that the calibration of the 24 $\mu$m-to-H$\alpha$ ratio to H$\alpha$ attenuation is not biased by any certain form of the attenuation curve.

In terms of the 24 $\mu$m-to-H$\alpha$ ratio, H$\alpha$ attenuation is expressed as follows:
\begin{equation}
A(\mathrm{H}\alpha) = 2.5 \log(1 + \frac{aL(24~\mu
\mathrm{m})}{L(\mathrm{H}\alpha)_\mathrm{obs}}) . \label{Eq_Att_Ha}
\end{equation}
We adopt this formula to estimate H$\alpha$ attenuation for the clusters inside the galaxies in our work, with the scaling factor of $a = 0.031 \pm 0.006$ according to the calibration for HII regions inside nearby galaxies addressed in \citet{2007ApJ...666..870C}.\footnote{Alternatively, \citet{2007ApJ...671..333K} have found a best-fitting value of $a = 0.038 \pm 0.005$ for HII regions inside a specific galaxy NGC~5194, which is comparable to the \citet{2007ApJ...666..870C} calibration, and adopting this value does not impose any effective bias on the results in this paper. \citet{2009ApJ...703.1672K} have derived $a = 0.020 \pm 0.005$ from integrated measurements of nearby galaxies, and this value is inappropriate for our work, because of the different stellar populations contained in the young clusters studied in this paper and the galaxies as a whole sampled in \citet{2009ApJ...703.1672K}.}

With H$\alpha$ attenuation derived as described above, we employ the combination of FUV and H$\alpha$ attenuations to probe the linear background slope. Figure \ref{IRHaUV_Att} shows 0.44$A$(H$\alpha$) as a function of $A$(FUV) for the galaxies in our sample, where 0.44$A$(H$\alpha$) is defined to convert the attenuation of the H$\alpha$ emission line to the attenuation of the stellar continuum at the H$\alpha$ band ($\sim 6563~\mathrm{Angstrom}$), and stellar continuum attenuation at H$\alpha$ band is defined as $A$(6563). The conversion $A$(6563) = 0.44$A$(H$\alpha$) is applied to our work on the basis of the definition $E(\mathrm{B}-\mathrm{V})_\mathrm{star} = 0.44 E(\mathrm{B}-\mathrm{V})_\mathrm{gas}$ in \citet{1997AIPC..408..403C} together with the assumption of a common attenuation law for stars and gas populating the same subregion inside one galaxy. The lines superimposed on each panel of this figure are reproduced by the MW curve (with the MW-type slope), the M-S curve (with the MW-type slope), the SMC curve (with the SMC-type slope), the S-M curve (with the SMC-type slope), and the C00 curve. These modeled lines indicate a clear separation between different slopes in the attenuation curve, and at the same time they present tight overlaps between the attenuation curves with the same slope albeit different bump strengths, e.g., the MW curve overlaps with the M-S curve, and the SMC curve overlaps with the S-M curve.\footnote{In Figure \ref{IRHaUV_Att}, we can see slight displacement in each overlap, which is due to the blueward tail of the 2175 Angstrom bump extending to the FUV bandpass. But compared with the separation, the displacements are very trivial and negligible.} As illustrated in this figure, the steeper slope leads to higher $A$(FUV) at constant $A$(H$\alpha$) compared with the shallower slope. This disparity between different slopes becomes more extensive as attenuation increases, and when 0.44$A$(H$\alpha$) $>$ 0.44 mag (i.e., $A$(H$\alpha$) $>$ 1.0 mag), the difference between $A$(FUV) reproduced by the MW- and the SMC-type slopes extends to over 1.0 mag.

In the top left panel of Figure \ref{IRHaUV_Att}, most of the data points from NGC~3031 lie in $A$(FUV) $<$ 2.0 mag and 0.44$A$(H$\alpha$) $<$ 0.3 mag. The low level of attenuation hampers strict constraints on the form of the attenuation curve for this galaxy. Nevertheless, the $A$(H$\alpha$) versus $A$(FUV) relation does not show an evident trend of the linear background being shallower than the MW-type slope, and in order to fulfill the $A$(FUV) versus $A(\mathrm{FUV})-A(\mathrm{NUV})$ distribution in Figure \ref{comp_Att}, the presence of a prominent 2175 Angstrom bump is assumed in the attenuation curve for NGC~3031.

\begin{figure*}[!ht]
\centering
\vspace*{-18mm}
\includegraphics[width=1.8\columnwidth]{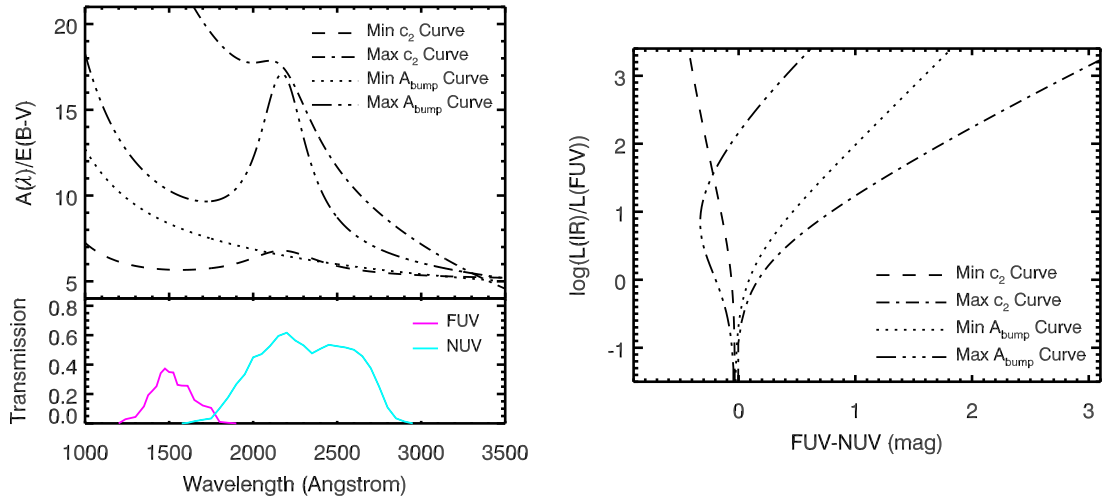}
\vspace*{-23mm}
\caption{Left: four extreme attenuation curves in the artificial sample with the minimal $c_2$ (dashed line; $c_2=0.100$, $c_3=1.949$, $c_4=0.107$, $\gamma=1.301$), the maximal $c_2$ (dot-dashed line; $c_2=5.000$, $c_3=3.971$, $0.349$, $\gamma=0.898$), the minimal $A_\mathrm{bump}$ (dotted line; $c_2=0.872$, $c_3=0.0$, $c_4=0.115$, $\gamma=0.830$), and the maximal $A_\mathrm{bump}$ (triple-dot-dashed line; $c_2=1.273$, $c_3=4.996$, $c_4=0.321$, $\gamma=0.503$). Filter transmission curves of the \emph{GALEX} FUV (magenta) and NUV (cyan) bands are also shown. Right: IRX-UV tracks reproduced with the attenuation curves shown in the left panel.}\label{mockCurv_IRXUV}
\end{figure*}

For NGC~4536, in the top right panel of Figure \ref{IRHaUV_Att}, the galactic core is located at $A$(FUV) $\sim 6.0$ mag and 0.44$A$(H$\alpha$) $\sim 1.8$ mag. Other data points from this galaxy span about 1.5 mag in
0.44$A$(H$\alpha$) within the range of $\sim 4.0$ mag in $A$(FUV), and they present a larger amount of H$\alpha$-band continuum attenuation at constant $A$(FUV) than the expectation from the MW-type slope (or the C00 curve). In this case, the locus appears to reflect the linear background shallower than the MW-type slope. However, the relation between $A$(FUV) and $A(\mathrm{FUV})-A(\mathrm{NUV})$ for NGC~4536 in Figure \ref{comp_Att} requires the MW-type slope with no bump, or steeper slopes with a certain degree of bumps in the attenuation curve. Otherwise, the linear background shallower than the MW-type slope has to result in a bump with negative strength, as can be deduced from the combination of Figures \ref{comp_Att} and \ref{IRHaUV_Att}. This discrepancy possibly comes from the estimates of H$\alpha$ attenuation, ascribed to differences in the conversion factor from 0.44 (which will be discussed in Section \ref{Sec_disc_Att_Ha_mock}), or/and different intrinsic values of 24 $\mu$m-to-H$\alpha$ ratio (which will be discussed in Section \ref{Sec_disc_NGC4536}), where the former potentially affects the results for all the galaxies in our work, and the latter is suspected to take place in NGC~4536. Despite the problem in the diagnosis of the linear background for NGC~4536, we do not find any evidence of the linear background in the attenuation curve steeper than the MW type, and also the suggestion of a bumpless feature in the attenuation curve is more apt for this galaxy.

The distribution of the clusters inside NGC~5194 approximates in the range of 0.0$-$4.0 mag in $A$(FUV) and 0.0$-$1.0 mag in 0.44$A$(H$\alpha$), which is well interpreted with the linear background slope between the MW and SMC types, as shown in the bottom left panel of Figure \ref{IRHaUV_Att}. This diagnosis in combination with the result from Figure \ref{comp_Att} strengthens the suggestion of the bumpless feature in the attenuation curve for this galaxy.

The correlation between $A$(H$\alpha$) and $A$(FUV) for NGC~7331 is shown in the bottom right panel of Figure \ref{IRHaUV_Att}. The disk clusters populate in $A$(FUV) $<$ 4.0 mag and 0.44$A$(H$\alpha$) $<$ 0.9 mag space, while the ring clusters are located at higher attenuation levels of about 3.0$-$6.3 mag in $A$(FUV) and 0.5$-$1.0 mag in 0.44$A$(H$\alpha$). A number of data points from the disk and the ring appear to possess a common regime between the MW- and SMC-type slopes. By taking the separate distributions between the disk and ring points in Figure \ref{comp_Att} into consideration, the ring clusters can be confirmed to have a bumpless feature (or a very trivial bump if any) in the attenuation curve, whereas the disk clusters are suggested as having a the similar slope but a prominent bump in order to better fit the relatively smaller differences between $A$(FUV) and $A$(NUV) than the ring clusters shown in Figure \ref{comp_Att}.

\section{INVESTIGATION WITH AN ARTIFICIAL SAMPLE}\label{Sec_mock}

In the above sections, the linear background and the 2175 Angstrom bump are expected to have significant impacts on the IRX-UV relation, and the observational features of the two parameters are diagnosed for the galaxies studied in this paper. In this section, by constructing an artificial sample to simulate observational data, we will investigate the observability of the presented features, assess the reliability of the applied diagnostics, and propose a prescription for quantitative constraints on the parameters in the attenuation curve.

\subsection{Artificial Sample}

We employ the models of the stellar population synthesis described in Section \ref{Sec_modeling} with a variety of attenuation curves to produce 20000 mock objects that compose a mock catalog. In this catalog, values for each physical parameter are randomly generated and uniformly distributed in a certain range, including stellar population age from 2 to 100 Myr, $0.0 < A(\mathrm{V}) \leq 3.0$ mag, $0.1 \leq c_2 \leq 5.0$, $0.0 \leq c_3 \leq 5.0$, $0.0 \leq c_4 \leq 0.5$, and $0.5 \leq \gamma \leq 1.5 ~\mu \mathrm{m}^{-1}$ \citep[see similar methods introduced in][]{2003MNRAS.341...33K, 2004MNRAS.349..769K, 2013ApJ...769..127L}.

\begin{figure*}[!ht]
\centering
\vspace*{-10mm}
\includegraphics[width=2.0\columnwidth]{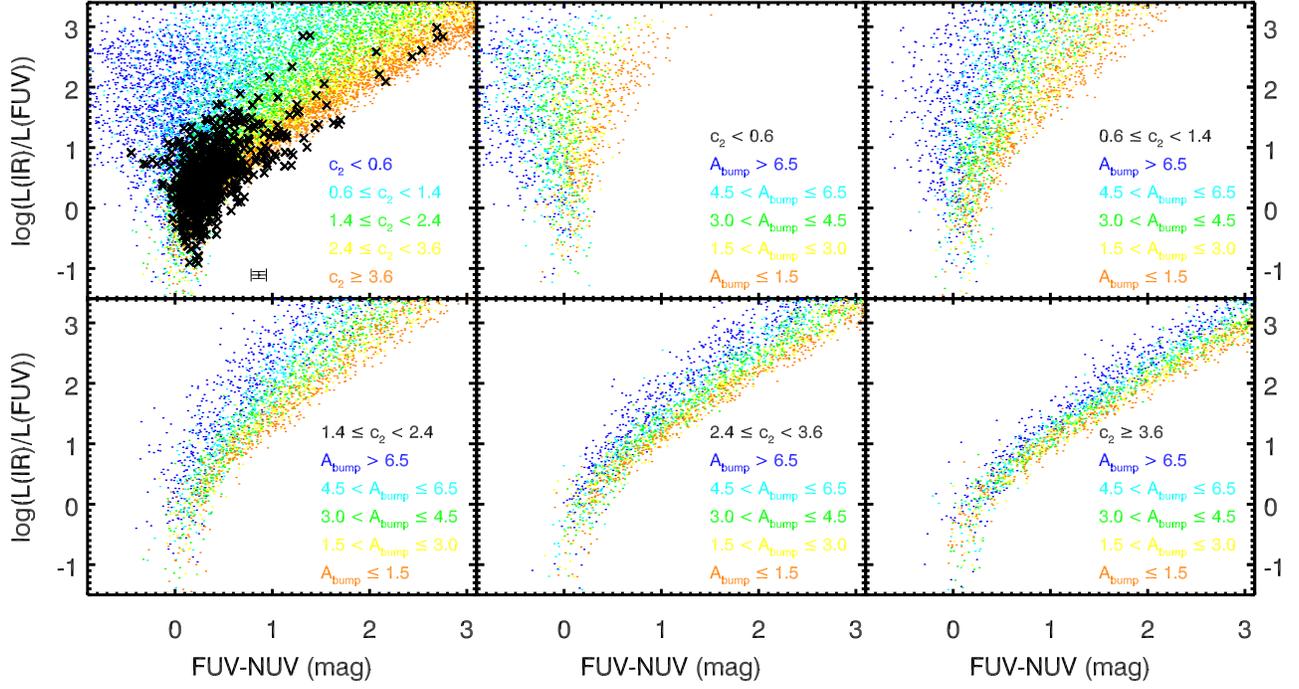}
\vspace*{-12mm}
\caption{IRX vs. $\mathrm{FUV}-\mathrm{NUV}$ for the artificial sample described in Section \ref{Sec_mock}. In the top left panel, the points in the entire sample are color-coded by $c_2$: $c_2 < 0.6$ (blue), $0.6 \leq c_2 < 1.4$ (cyan), $1.4 \leq c_2 < 2.4$ (green), $2.4 \leq c_2 < 3.6$ (yellow), and $c_2 \geq 3.6$ (orange); black crosses superimposed on this panel represent all the clusters inside the four galaxies described in Section \ref{Sec_results}; and the error bar plotted at the bottom of this panel shows the median uncertainties for the clusters. The other five panels show IRX as a function of $\mathrm{FUV}-\mathrm{NUV}$ for the subsample with $c_2 < 0.6$ (top middle panel, corresponding to the blue points in the top left panel), $0.6 \leq c_2 < 1.4$ (top right panel, corresponding to the cyan points in the top left panel), $1.4 \leq c_2 < 2.4$ (bottom left panel, corresponding to the green points in the top left panel), $2.4 \leq c_2 < 3.6$ (bottom middle panel, corresponding to the yellow points in the top left panel), and $c_2 \geq 3.6$ (bottom right panel, corresponding to the orange points in the top left panel). The points in each subsample are color-coded by $A_\mathrm{bump}$: $A_\mathrm{bump} > 6.5$ (blue), $4.5 < A_\mathrm{bump} \leq 6.5$ (cyan), $3.0 < A_\mathrm{bump} \leq 4.5$ (green), $1.5 < A_\mathrm{bump} \leq 3.0$ (yellow), and $A_\mathrm{bump} \leq 1.5$ (orange).} \label{IRXUV2_mock}
\end{figure*}

This artificial sample, for the purpose of reproducing attenuation features that potentially exist in young clusters inside galaxies, only contains young and simple stellar populations ($\leq 100$ Myr); older or more composite stellar populations are not adopted. The $A$(V) up to 3.0 mag suffices to fit most observations. The ranges for the coefficients in the FM parameterization of the attenuation curve, i.e., $c_2$, $c_3$, $c_4$, and $\gamma$, are determined in consideration of the realistic observations \citep{1988ApJ...328..734F, 1990ApJS...72..163F, 2003ApJ...594..279G}. The stellar populations in this catalog are assumed to be born from an instantaneous burst with the \citet{2002Sci...295...82K} initial mass function (with exponents of 1.3 over 0.1$-$0.5 $M_{\odot}$ and 2.3 over 0.5$-$100 $M_{\odot}$) and solar metallicity.

UV and IR luminosities for the mock objects are obtained, respectively, through the convolution of the modeled spectra with filter transmission curves and the sum of attenuated stellar emission, which is the same as
described in Section \ref{Sec_modeling}. Uncertainties are also simulated and assigned to the quantities in the artificial sample, and at this step, we assume a normal distribution with the standard deviation defined as the median values of the photometric uncertainties in the observational data described in Section \ref{Sec_data}.

The investigation with the mock catalog will focus on $c_2$ and $A_{\mathrm{bump}}$ ($\equiv \pi c_3 /(2 \gamma)$, the area enclosed by the 2175 Angstrom bump in the attenuation curve) and involve an elaborate analysis of the roles played by both parameters in the IRX-UV relation. Figure \ref{mockCurv_IRXUV} is displayed as a preview to show the extreme attenuation curves with minimal and maximal $c_2$ and $A_{\mathrm{bump}}$ in the artificial sample respectively, and the IRX-UV tracks reproduced with these attenuation curves. The following subsections will present the investigation with all the attenuation curves of high diversity in the artificial sample.

\subsection{Simulated IRX-UV Relation}

Figure \ref{IRXUV2_mock} shows the IRX-UV diagrams for the artificial sample and its different ingredients. The top left panel of this figure presents the result for the entire sample, where the mock objects are color-coded by different $c_2$ ranges, and the observational data (the clusters inside the galaxies) are superimposed on the diagram in order to show the artificial sample covering the observations in the IRX-UV plane; the other five panels exhibit the results for the five subsamples of the mock objects that correspond, respectively, to the five $c_2$ bins defined in the top left panel, where the data points are color-coded by different $A_{\mathrm{bump}}$ ranges.

In the top left panel of Figure \ref{IRXUV2_mock}, the entire sample
spreads in a broad range in the IRX-UV diagram and possesses more
extended $\mathrm{FUV}-\mathrm{NUV}$ space with increasing IRX. By means of the color-coding, we can see a clear trend of the overall data distribution with $c_2$: the objects with larger $c_2$ tend to incline toward redder
$\mathrm{FUV}-\mathrm{NUV}$ at fixed IRX. However, this trend presents
considerable dispersion on local scales, where the loci for different $c_2$ bins overlap seriously with each other. This dispersion is interpreted in the other panels of Figure \ref{IRXUV2_mock} by further categorizing each of the subsamples into different $A_\mathrm{bump}$ bins: in addition to $c_2$, $A_\mathrm{bump}$ affects the IRX-UV relation at the same time but in an opposite direction, displayed as a blueing of the UV color at constant IRX. These results clearly illustrate that both of $c_2$ and $A_\mathrm{bump}$ are able to have significant impacts on the locations of data points in the IRX-UV planes, whereas the degeneracy of the two parameters, in turn, is likely to disperse the visible signatures of any single parameter.

\subsection{Constraints on the Attenuation Curve}\label{Sec_conCurv}

\begin{figure*}[!ht]
\centering
\vspace*{-12mm}
\includegraphics[width=1.8\columnwidth]{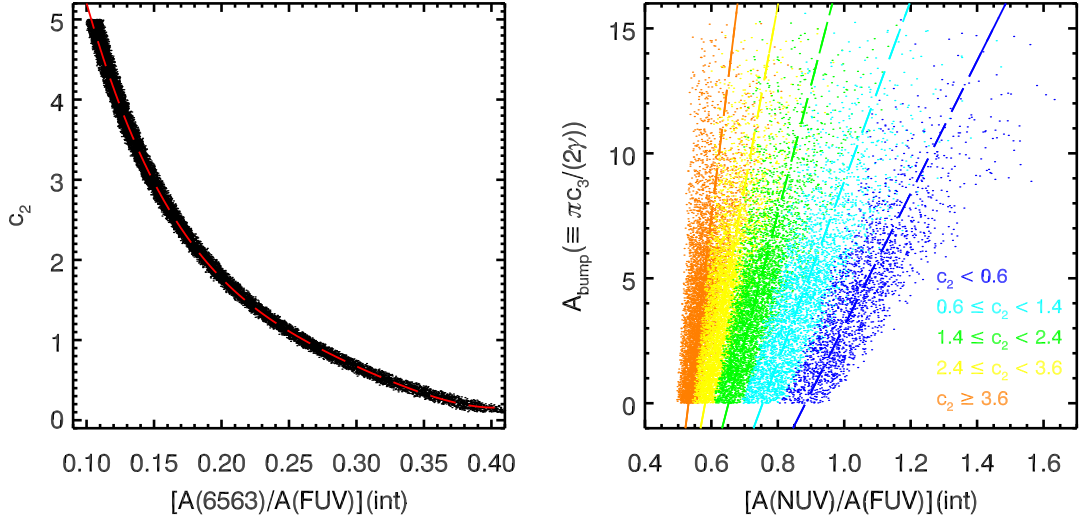}
\vspace*{-14mm}
\caption{Left: $c_2$ vs. ratio of intrinsic 6563 Angstrom to FUV attenuations for the artificial sample described in Section \ref{Sec_mock}, and the red dashed line is the best-fitting curve for the data, defined as Equation (\ref{Eq_c2}). Right: $A_\mathrm{bump}$ vs. ratio of intrinsic NUV to FUV attenuations for the same artificial sample, where the points in this panel are color-coded by $c_2$: $c_2 < 0.6$ (blue), $0.6 \leq c_2 < 1.4$ (cyan), $1.4 \leq c_2 < 2.4$ (green), $2.4 \leq c_2 < 3.6$ (yellow), and $c_2 \geq 3.6$ (orange), and the dashed lines show the best-fitting curves for the subsamples with identical colors.} \label{IRHaUV_Att_mock}
\end{figure*}

\begin{figure*}[!ht]
\centering
\vspace*{-12mm}
\includegraphics[width=1.8\columnwidth]{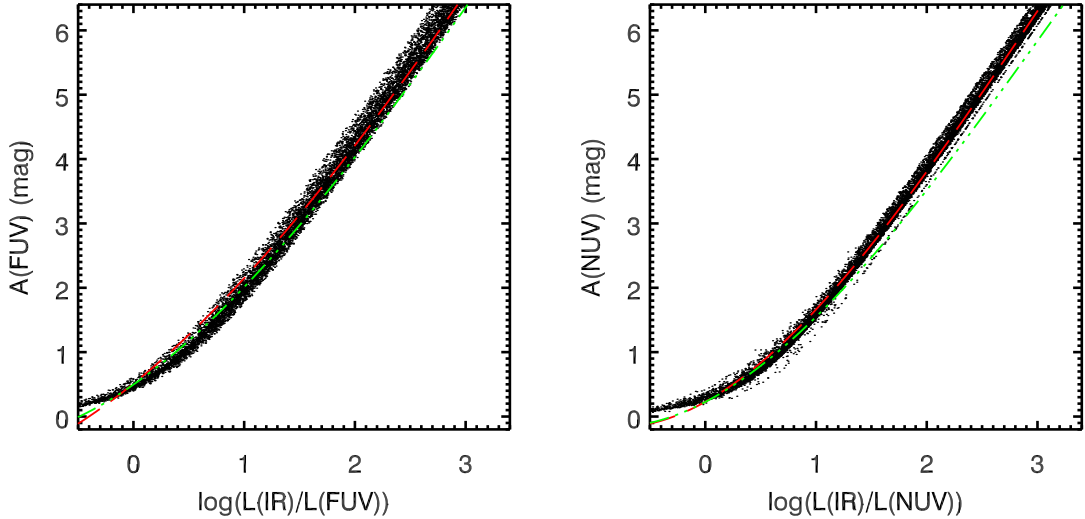}
\vspace*{-14mm}
\caption{$A$(FUV) as a function of
$\log(L(\mathrm{IR})/L(\mathrm{FUV}))$ in the left panel, and
$A$(NUV) as a function of $\log(L(\mathrm{IR})/L(\mathrm{NUV}))$ in
the right panel, for the artificial sample described in Section
\ref{Sec_mock}. In both panels, the red dashed lines are is the best
fitting curves for the data, and the green triple-dot-dashed lines
show the B05 relations defined as Equations (\ref{Eq_Buat05a}) and
(\ref{Eq_Buat05b}).
\vspace*{6mm}
} \label{AttIRX_mock}
\end{figure*}

The combination of FUV, NUV, and H$\alpha$-band continuum attenuations is a feasible way to estimate the linear background slope and the 2175 Angstrom bump strength, as suggested in Section \ref{Sec_ExtCurv}. In this subsection, we make use of the artificial sample to thoroughly investigate the correlations of $c_2$ and $A_\mathrm{bump}$ with $A$(FUV), $A$(NUV), and H$\alpha$-band continuum attenuation (i.e., $A$(6563)),\footnote{With regard to the conversion $A$(6563) = 0.44$A$(H$\alpha$) applied in this paper, we will discuss the uncertainties introduced by variations in the conversion factor from 0.44 in the results in Section \ref{Sec_disc_Att_Ha_mock}.} for the sake of quantitative constraints on the parameters in the attenuation curve from observational properties.

Figure \ref{IRHaUV_Att_mock} shows $c_2$ as a function of $A$(6563)/$A$(FUV) and $A_\mathrm{bump}$ as a function of $A$(NUV)/$A$(FUV) for the artificial sample. In the left panel, $A$(6563)/$A$(FUV) correlates tightly with $c_2$, demonstrating that $A$(6563)/$A$(FUV) is in a good position to trace the linear background slope. In the right panel, the dependence of $A$(NUV)/$A$(FUV) on $A_\mathrm{bump}$ is almost invisible, but with a series of constant $c_2$ ranges assigned, the correlation between $A$(NUV)/$A$(FUV), $A_\mathrm{bump}$, and $c_2$ can be obviously seen. This figure serves as a vivid description of the degeneracy between the linear background slope and the 2175 Angstrom bump in \emph{GALEX} UV bands and at the same time manifests the validity of the combination of $A$(FUV), $A$(NUV), and $A$(6563) in estimating $c_2$ and $A_\mathrm{bump}$. In order to quantify the estimation of $c_2$ and $A_\mathrm{bump}$ from $A$(FUV), $A$(NUV), and $A$(6563), we conduct polynomial fits for the two correlations. The best-fitting formulae are displayed as follows:
\begin{equation}
c_2 = 1263.35 x_3^4 - 1555.11 x_3^3 + 733.92 x_3^2 - 164.59 x_3 + 15.75,
\label{Eq_c2}
\end{equation}
and
\begin{equation}
A_\mathrm{bump} = 0.73 x_3^2 x_4 - 100.78 x_3 x_4 + 56.19 x_4 + 215.99 x_3^2 - 86.42 x_3 \\
 - 13.94,
\label{Eq_Abump}
\end{equation}
where $x_3 \equiv A(6563)/A(\mathrm{FUV})$ and $x_4 \equiv
A(\mathrm{NUV})/A(\mathrm{FUV})$. Because of the upper and lower limits
in the parameter ranges in the fitting, we suggest that Equations (\ref{Eq_c2}) and (\ref{Eq_Abump}) are applicable to the data within the range of $0.1 \leq A(6563)/A(\mathrm{FUV}) \leq 0.4$ and $0.4 \leq A(\mathrm{NUV})/A(\mathrm{FUV}) \leq 1.6$. Uncertainties in the estimates of $c_2$ and $A_\mathrm{bump}$ from $A$(FUV), $A$(NUV), and $A$(6563) by adopting Equations (\ref{Eq_c2}) and (\ref{Eq_Abump}) will be discussed in the paragraphs below.

In the previous section of this paper, $A$(FUV) and $A$(NUV) for the observational data are derived from $\log(L(\mathrm{IR})/L(\mathrm{FUV}))$ and $\log(L(\mathrm{IR})/L(\mathrm{NUV}))$, respectively, by using the B05
calibration, and then they are adopted to probe the features of the attenuation curve. In this situation, it is necessary to examine the reliability of the B05 calibration, i.e., whether or not the B05 calibration is independent of any presumption of the attenuation curve.

\begin{figure}[!ht]
\centering
\hspace*{-14mm}
\includegraphics[width=1.4\columnwidth]{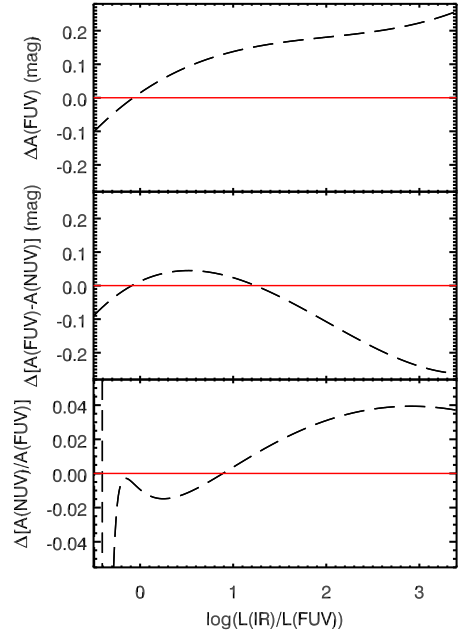}
\caption{Top: margin between $A$(FUV) estimated with the
calibrations in this work (red dashed line in Figure
\ref{AttIRX_mock}) and in B05 (green triple-dot-dashed line in
Figure \ref{AttIRX_mock}), as a function of
$\log(L(\mathrm{IR})/L(\mathrm{FUV}))$. Bottom: margin between
$A$(NUV)/$A$(FUV) estimated with the calibrations in this work (red
dashed line in Figure \ref{AttIRX_mock}) and in B05 (green
triple-dot-dashed line in Figure \ref{AttIRX_mock}), as a function
of $\log(L(\mathrm{IR})/L(\mathrm{FUV}))$.} \label{Del_AttIRX_mock}
\end{figure}

Our mock objects cover a wide variety of attenuation curves and therefore serve as an excellent sample for this examination. In Figure \ref{AttIRX_mock}, we plot $A$(FUV) as a function of $\log(L(\mathrm{IR})/L(\mathrm{FUV}))$ and $A$(NUV) as a function of $\log(L(\mathrm{IR})/L(\mathrm{NUV}))$ for the artificial sample, and we also superimpose the best-fitting curves for our mock data as well as the B05 calibration curves on both diagrams for comparison purposes. The tight locus composed by the data points in each panel of this figure proves the IR-to-UV ratios to be intrinsically insensitive to the shapes of the attenuation curve. The best-fitting curves for the mock objects are formulated as the following equations:
\begin{equation}
A(\mathrm{FUV}) = - 0.0200 x_1^3 + 0.2722 x_1^2 + 1.3859 x_1 + 0.5112 , \label{Eq_ourfit_a}
\end{equation}
and
\begin{equation}
A(\mathrm{NUV}) = - 0.0597 x_2^3 + 0.5334 x_2^2 + 0.9605 x_2 + 0.2287 , \label{Eq_ourfit_b}
\end{equation}
where $x_1 \equiv \log(L(\mathrm{IR})/L(\mathrm{FUV}))$ and $x_2 \equiv \log(L(\mathrm{IR})/L(\mathrm{NUV}))$. The B05 calibration is in good agreement with our fitting in each panel of Figure \ref{AttIRX_mock}. The maximum difference between the two calibrations is 0.25 mag in either $A$(FUV) or $A(\mathrm{FUV})-A(\mathrm{NUV})$ for $-0.5 < \mathrm{IRX} < 3.4$, and 0.04 in $A$(NUV)/$A$(FUV) for $-0.2 < \mathrm{IRX} < 3.4$, as shown in Figure \ref{Del_AttIRX_mock}. This examination implies that even if we change the formula during the estimation of FUV and NUV attenuations, there will not be any significant effect on the results; a remarkable disparity between the two calibrations occurs only when IRX $> 4.0$, which is beyond most observations. Consequently, the B05 calibration in general cases offers a reliable, unbiased diagnosis of the attenuation curve.

\begin{figure}[!ht]
\centering
\hspace*{-14mm}
\includegraphics[width=1.4\columnwidth]{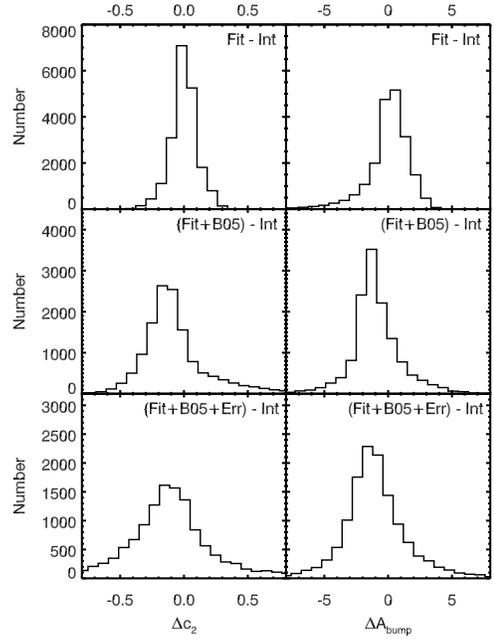}
\caption{Histograms of $\Delta c_2$ (left column) and $\Delta
A_\mathrm{bump}$ (right column) for the artificial sample described in
Section \ref{Sec_mock}. $\Delta c_2$ and $\Delta A_\mathrm{bump}$
are the differences in $c_2$ and $A_\mathrm{bump}$ between the
estimated and intrinsic values described in Section
\ref{Sec_conCurv}. All the data in the mock catalog are plotted in the top row, while the data with IRX $\leq 4.0$ are shown in the middle and bottom rows.} \label{comp_Att6_mock}
\end{figure}

\begin{figure}[!ht]
\centering
\hspace*{-14mm}
\includegraphics[width=1.4\columnwidth]{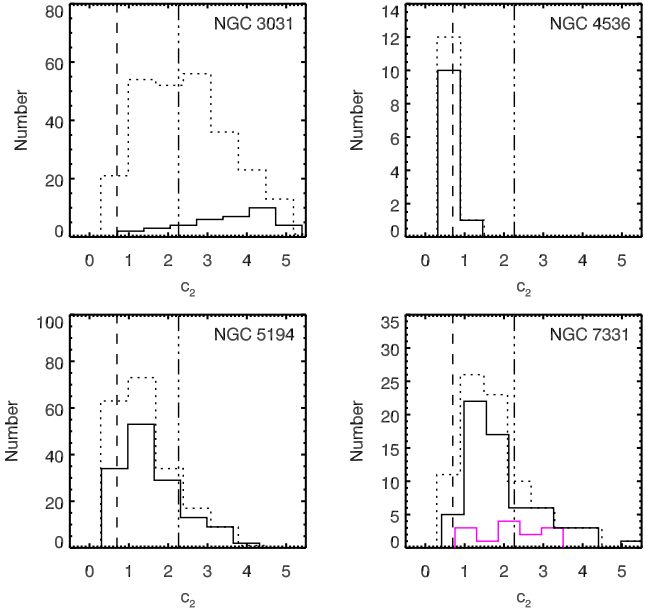}
\caption{Histograms of $c_2$ for the UV and IR clusters inside NGC~3031 (top left panel), NGC~4536 (top right panel), NGC~5194 (bottom left panel), and NGC~7331 (bottom right panel), estimated by adopting Equation (\ref{Eq_c2}). In each panel, the solid line describes the clusters with IRX $> 0.6$, while the dotted line shows all the clusters measured in the galaxy; specifically for NGC~7331, the black lines represent the disk clusters, and the magenta line represents the ring clusters. The dashed line marks the value of the MW $c_2$, and the triple-dot-dashed line marks the value of the SMC $c_2$.} \label{hist_c2}
\end{figure}

The artificial sample also helps to assess uncertainties in the prescription for constraints on the attenuation curve. In the top panel of Figure \ref{comp_Att6_mock}, estimates of $c_2$ and $A_\mathrm{bump}$ from intrinsic $A$(FUV) and $A$(NUV) by adopting Equations (\ref{Eq_c2}) and (\ref{Eq_Abump}) have a standard deviation of $\sim \pm 0.1$ in $c_2$ and $\sim \pm 1.7$ in $A_\mathrm{bump}$. With the B05 calibration used to obtain $A$(FUV) and $A$(NUV) from $\log(L(\mathrm{IR})/L(\mathrm{FUV}))$ and
$\log(L(\mathrm{IR})/L(\mathrm{NUV}))$ but without photometric uncertainties imposed on the quantities, the application of Equations (\ref{Eq_c2}) and (\ref{Eq_Abump}) introduces a slight offset of $\sim -0.1$ in $c_2$ and $\sim -1.2$ in $A_\mathrm{bump}$ and a standard deviation of $\sim \pm 0.3$ in $c_2$ and $\sim \pm 2.2$ in $A_\mathrm{bump}$, which can be seen from the middle panels of this figure.\footnote{In the middle and bottom panels of Figure \ref{comp_Att6_mock}, we do not plot the data in the artificial sample with IRX $> 4.0$, in order to avoid the additional error resulting from the application of the B05 formula to IRX $> 4.0$, as mentioned in the above paragraph.} After adding the simulated photometric uncertainties to the mock data, Equations (\ref{Eq_c2}) and (\ref{Eq_Abump}) in combination with the B05 formula result in the same level of offset and the standard deviation of $\sim \pm 0.5$ in $c_2$ and $\sim \pm 2.8$ in $A_\mathrm{bump}$, as shown in the bottom panels in Figure \ref{comp_Att6_mock}. This examination reflects that uncertainties introduced by calibration and photometry are likely to deviate the estimates of the parameters in the attenuation curve from the intrinsic values at a certain degree, but effective constraints can still be made with the uncertainties taken into account and are able to at least distinguish between the MW- and SMC-type signatures.

\begin{figure}[!ht]
\centering
\hspace*{-14mm}
\includegraphics[width=1.4\columnwidth]{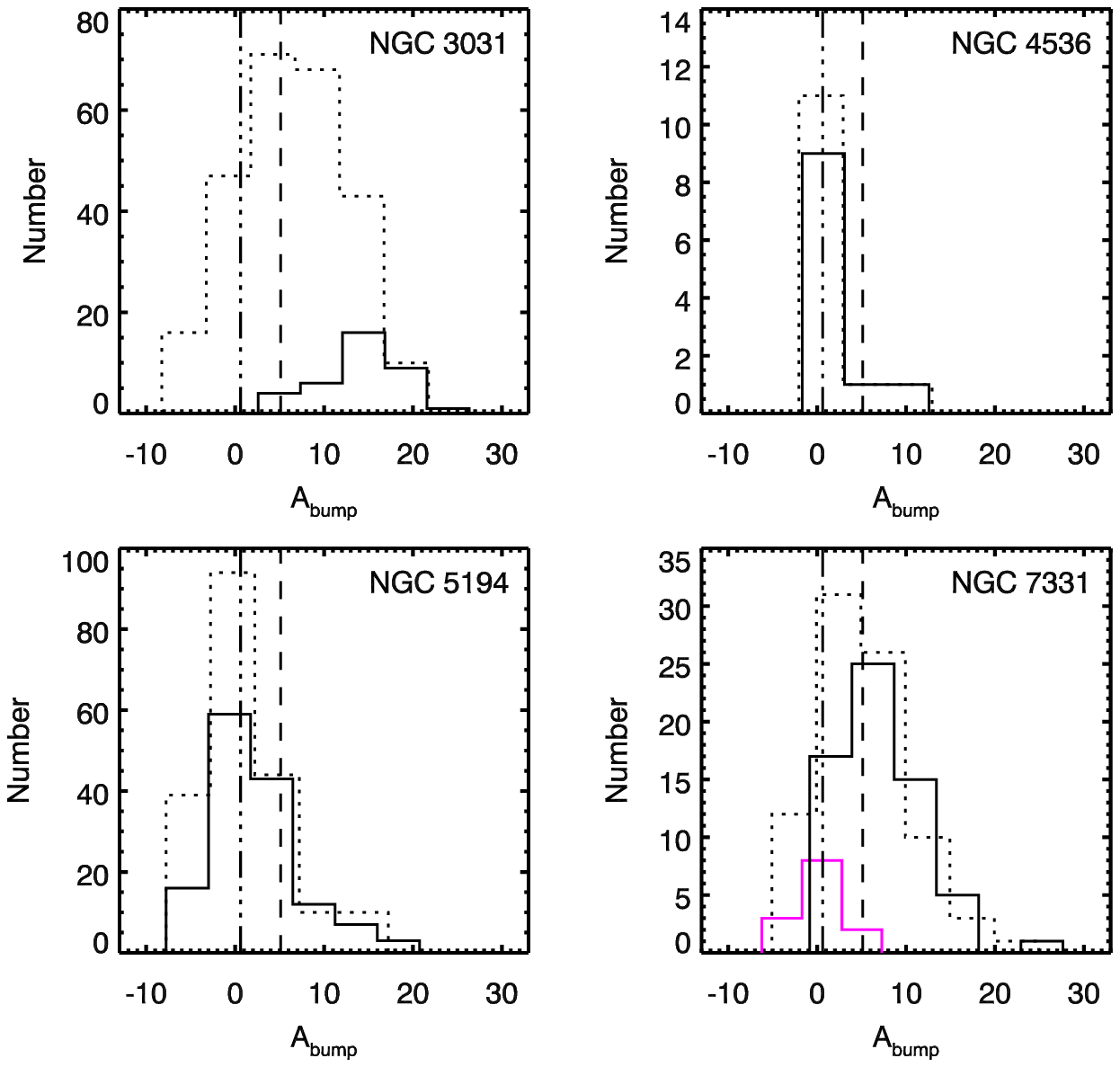}
\caption{Histograms of $A_\mathrm{bump}$ for the UV and IR clusters inside NGC~3031 (top left panel), NGC~4536 (top right panel), NGC~5194 (bottom left panel), and NGC~7331 (bottom right panel) estimated by adopting Equation (\ref{Eq_Abump}). In each panel, the solid line describes the clusters with IRX $> 0.6$, while the dotted line shows all the clusters measured in the galaxy; specifically for NGC~7331, the black lines represent the disk clusters, and the magenta line represents the ring clusters. The dashed line marks the value of the MW $A_\mathrm{bump}$, and the triple-dot-dashed line marks the value of the SMC $A_\mathrm{bump}$.} \label{hist_Abump}
\end{figure}

In addition to the B05 formula, \citet{2005MNRAS.360.1413B} offers similar formulae to estimate $A$(FUV) and $A$(NUV) from $\log(L(\mathrm{IR})/L(\mathrm{FUV}))$ and
$\log(L(\mathrm{IR})/L(\mathrm{NUV}))$. By comparing this calibration with the B05 formula, we find very trivial differences between the results, with a mean value of $< 0.03$ mag in $A$(FUV) and $\sim 0.05$ mag in $A(\mathrm{FUV})-A(\mathrm{NUV})$. Also, there are several other formulae for the conversion from $\log(L(\mathrm{IR})/L(\mathrm{FUV}))$ to $A$(FUV)
\citep[e.g.,][]{1999ApJ...521...64M, 2004MNRAS.349..769K,
2011ApJ...741..124H, 2012A&A...539A.145B}. As discussed elaborately in \citet{2011ApJ...741..124H, 2012A&A...539A.145B}, these various estimates of $A$(FUV), including those with the B05 calibration, are very close to each other, but in most of the formulas, the estimate of $A$(NUV) requires a presumption of a certain type of attenuation curve. In this case, the combination of $A$(FUV) and $A$(NUV) is inherently biased and therefore fails to trace the features of the attenuation curve. Because of this factor, we employ the B05 formula, which separately calibrates $A$(FUV) and $A$(NUV) in our work.

With the application of Equations (\ref{Eq_c2}) and (\ref{Eq_Abump}) to the clusters (with $0.1 \leq 0.44A(\mathrm{H}\alpha)/A(\mathrm{FUV}) \leq 0.4$, and $0.4 \leq A(\mathrm{NUV})/A(\mathrm{FUV}) \leq 1.6$) inside NGC~3031, NGC~4536, NGC~5194, and NGC~7331, we estimate $c_2$ and $A_\mathrm{bump}$ from $A$(FUV), $A$(NUV), and $A$(H$\alpha$) for these galaxies. Figures \ref{hist_c2} and \ref{hist_Abump} show the histograms of the derived $c_2$ and $A_\mathrm{bump}$ for each of the galaxies, where we plot the clusters with IRX $> 0.6$ as the solid line and a total number of the clusters as the dotted line, since the estimates of $c_2$ and $A_\mathrm{bump}$ for objects with IRX $< 0.6$ are potentially influenced by variations in stellar population age, which will be discussed in Section \ref{Sec_disc_age2}.

In Figures \ref{hist_c2} and \ref{hist_Abump}, taking the total number of the clusters into account, NGC~3031 presents the linear background slope spanning over a broad range with a typical value of about 2.2 in $c_2$, comparable to the SMC slope, and the 2175 Angstrom bump stronger than
the MW type with $A_\mathrm{bump} > 5.0$ for a majority of the clusters. On the contrary, the attenuation curve for NGC~4536 appears with shallow and bumpless features ($c_2 \sim 0.6$ and $A_\mathrm{bump} \sim 0.0$) and can be characterized by the MW-type linear background but with no bump.\footnote{Due to the criterion of $0.1 \leq 0.44A(\mathrm{H}\alpha)/A(\mathrm{FUV}) \leq 0.4$, a number of the clusters with high H$\alpha$ attenuation inside NGC~4536 are excluded in the estimation of $c_2$ and $A_\mathrm{bump}$.} The similar bumpless feature is displayed in NGC~5194, with the moderate linear background ranging between the MW- and the SMC-type slopes with a typical value of $c_2 \sim 1.3$ suggested in the attenuation curve. For NGC~7331, most of the disk clusters inside NGC~7331 present a moderate linear background slope comparable to that for NGC~5194 and a 2175 Angstrom bump as strong as the MW-type bump with $A_\mathrm{bump} \sim 5.0$ on average; the ring clusters cover a wide range in the linear background slope but appear to have a very trivial or even absent bump in the attenuation curve, similar to those for NGC~4536 and NGC~5194. The classification of the clusters with IRX $> 0.6$ does not lead to any obvious difference in the results except for NGC~3031 where only a small percentage of the clusters have IRX $> 0.6$ in contrast to the other galaxies. Compared with the majority of clusters in NGC~3031, the clusters with IRX $> 0.6$ are exhibited to have a steeper linear background and a stronger 2175 Angstrom bump.

Although Equations (\ref{Eq_c2}) and (\ref{Eq_Abump}) offer a prescription for quantitative constraints on the parameters in the attenuation curve, the combination of $A$(FUV), $A$(NUV), and $A$(H$\alpha$) remains to be a rough tracer. In any case, more accurate estimates of the attenuation curve are based on the analysis of the spectral continuum taken from spectroscopic observations.

\section{DISCUSSION}\label{Sec_disc}

Ever since the deviation in the IRX-UV relation was initially found in observations of galaxies, a number of studies $\textemdash$ from the interpretation offered by \citet{2004MNRAS.349..769K} a decade ago to the very recent research taken by \citet{2013ApJ...773..174G} $\textemdash$ have suggested stellar population age to be the second parameter in the IRX-UV function. Through a census of the impacts of various physical parameters on the IRX-UV relation conducted by the SED fitting technique, \citet{2012A&A...539A.145B} have suggested that, in addition to the intrinsic UV color of a stellar population being found to primarily determine the IRX-UV distribution for galaxies, a secondary but non-negligible role is played by the shape of the attenuation curve. The above sections in this paper present the influences of the attenuation curve on the IRX-UV relation; however, until now there has not been any successful observation of the features of the attenuation curve in the IRX-UV relation. As we have analyzed in the above section, the degeneracy of the linear background slope and the 2175 Angstrom bump strength in UV wavelength bands is likely to be one significant obstacle. In this section, we will discuss other possible reasons for the failure in observational measurements of the parameters related to the attenuation curve in the IRX-UV function.

\subsection{Age Effects on the IRX-UV Relation}\label{Sec_disc_age1}

\begin{figure}[!ht]
\centering
\hspace*{-14mm}
\includegraphics[width=1.4\columnwidth]{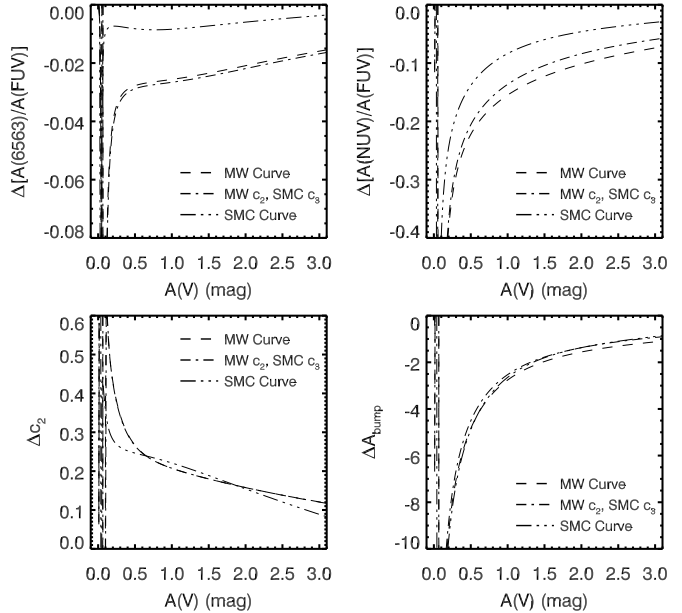}
\caption{Differences between 2 and 100 Myr stellar populations in $A$(6563)/$A$(FUV) (top left panel), $A$(NUV)/$A$(FUV) (top right panel), $c_2$ (bottom left panel), and $A_\mathrm{bump}$ (bottom right panel) as a function of $A$(V). The lines in each panel are obtained by spectral synthesis modeling with different attenuation curves: the MW curve (dashed line), the M-S curve (dot-dashed line), and the SMC curve (triple-dot-dashed line), which is described in Section \ref{Sec_modeling}. $A$(FUV) and $A$(NUV) are derived from $\log(L(\mathrm{IR})/L(\mathrm{FUV}))$ and
$\log(L(\mathrm{IR})/L(\mathrm{NUV}))$ by using Equations (\ref{Eq_Buat05a}) and (\ref{Eq_Buat05b}), and $c_2$ and $A_\mathrm{bump}$ are derived from the combination of $A$(FUV), $A$(NUV), and $A$(6563) by using Equations (\ref{Eq_c2}) and (\ref{Eq_Abump}).}
\label{check1_age_eff}
\end{figure}

As \citet{2012A&A...539A.145B} have pointed out, variations in intrinsic UV color are the main cause of the deviation in the IRX-UV relation. In most cases, stellar population age, compared with other physical parameters such as initial mass function and metallicity, dominates intrinsic UV color more effectively. In Paper I, we have investigated the influences of stellar population age on the IRX-UV relation, which appear to be a considerable offset toward redder UV colors for older stellar populations but appear to be gradually invisible as the star formation timescale increases. The maximum effect of stellar population age exists in simple stellar populations with the SFH of an instantaneous burst or a short star formation timescale (e.g., $\tau_\mathrm{SF} \leq 0.01$ Gyr). In this kind of SFH scenario, as shown in Figure 16 in Paper I as an example, the intrinsic UV color for a 100 Myr stellar population is redder than that for a 2 Myr stellar population by a factor of $\sim 0.5$ mag in $\mathrm{FUV}-\mathrm{NUV}$; as stellar populations continue to evolve, the offset in intrinsic UV color from the 2 Myr stellar population extends more intensively, for instance, to $\sim 2.0$ mag for a 500 Myr stellar population and to $\sim 4.0$ mag for a 1 Gyr stellar population. Even though more complex SFHs give birth to composite stellar populations and thus complicate the correlation between stellar population age and intrinsic UV color, there is no doubt that age effects substantially deviate the IRX-UV relation, particularly for systems containing evolved stellar populations.

In this paper, as can be seen from the right panel of Figure \ref{ExtCurve}, the difference between $\mathrm{FUV}-\mathrm{NUV}$ produced by the MW- and SMC-type attenuation curves for the same stellar population age (i.e., with the same intrinsic UV color) increases with rising IRX, for instance, to $\sim 0.5$ mag at IRX $\sim 0.5$ and to $\sim 2.0$ mag at IRX $\sim 2.5$. By combining the results in Paper I and those in this paper, we can imply that because of the age effects that seriously vary intrinsic UV color, the role played by the attenuation law in the IRX-UV relation is possibly obscured and therefore difficult to detect even though variations in the attenuation curve still introduce uncertainties into observational properties for galaxies. In this situation, the features of the attenuation law are distinguishable only for young (approximately $< 100$ Myr) and dust-rich (at lease IRX $> 0.5$) stellar populations; otherwise, stellar population age tends to have a predominant influence on the IRX-UV relation, as \citet{2013ApJ...773..174G} have confirmed by compiling and investigating a sample of galaxies with the stellar population age in the range of 10 Myr$-$5 Gyr and IRX $\leq 0.5$.

\begin{figure}[!ht]
\centering
\vspace*{-10mm}
\hspace*{-10mm}
\includegraphics[width=1.4\columnwidth]{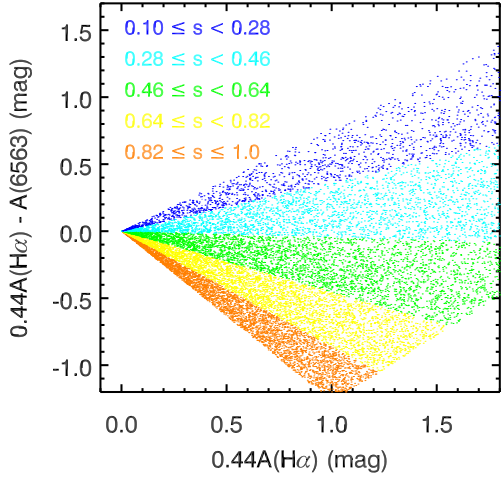}
\vspace*{-9mm}
\caption{Residuals between $0.44A(\mathrm{H}\alpha)$ and $A$(6563)
as a function of $0.44A(\mathrm{H}\alpha)$ for the artificial sample
described in Section \ref{Sec_disc_Att_Ha_mock}, where
$0.44A(\mathrm{H}\alpha) = 0.44(A(6563)/s)$. The points are
color-coded by the conversion factor $s$: $0.30 \leq s < 0.34$
(blue), $0.34 \leq s < 0.38$ (cyan), $0.38 \leq s < 0.42$ (green),
$0.42 \leq s < 0.46$ (yellow), and $0.46 \leq s < 0.50$ (orange).}
\label{check_Att_Ha_mock}
\end{figure}

\subsection{Requirement for Dust Content in the Diagnostics of the Attenuation Law}\label{Sec_disc_age2}

Accurate measurements of the parameters related to the attenuation curve are always a challenge. Without available spectral observations, a comparison between attenuations at different wavelength bands appears to be a feasible approach. In this paper, we adopt FUV, NUV, and 6563 Angstrom attenuations to probe the linear background and the 2175 Angstrom bump in the attenuation curve. The estimation of FUV and NUV attenuations is based on the B05 formula (Equations (\ref{Eq_Buat05a}) and (\ref{Eq_Buat05b}) deriving $A$(FUV) and $A$(NUV) from $\log(L(\mathrm{IR})/L(\mathrm{FUV}))$ and
$\log(L(\mathrm{IR})/L(\mathrm{NUV}))$, respectively. However, as \citet{2008MNRAS.386.1157C} have suggested, the correlation between $A$(UV) and $\log(L(\mathrm{IR})/L(\mathrm{UV}))$ is also biased by the stellar population age. Fortunately, the sample studied in our work contains only young clusters inside galaxies, and from the modeled lines in Figure \ref{IRXUV_gal} we can see that $\log(L(\mathrm{IR})/L(\mathrm{FUV}))$ varies by a factor of $< 0.2$ for the same amount of attenuation as long as stellar populations are younger than 100 Myr. Nevertheless, although this variation is too trivial to introduce deviation into the IRX-UV relation, the features of the attenuation curve probed by comparing attenuations at different bands are still unable to withstand age or/and other effects if dust content is low.

In order to explore the requirement for dust content in the diagnostics of the attenuation curve, we examine the systematic uncertainties in $A$(6563)/$A$(FUV), $A$(NUV)/$A$(FUV), $c_2$, and $A_\mathrm{bump}$ induced by the difference between 2 and 100 Myr stellar population ages for the three typical attenuation curves, the MW curve, the M-S curve, and the SMC curve, as a function of $A$(V), on the basis of the stellar population synthesis modeling described in Section \ref{Sec_modeling}. In this examination, $A$(FUV) and $A$(NUV) are derived from $\log(L(\mathrm{IR})/L(\mathrm{FUV}))$ and
$\log(L(\mathrm{IR})/L(\mathrm{NUV}))$ by using Equations (\ref{Eq_Buat05a}) and (\ref{Eq_Buat05b}), and $c_2$ and $A_\mathrm{bump}$ are derived from the combination of $A$(FUV), $A$(NUV), and $A$(6563) by using Equations (\ref{Eq_c2}) and (\ref{Eq_Abump}). The results are shown in Figure \ref{check1_age_eff}, where the difference in each of the compared quantities between 2 and 100 Myr stellar populations ($\mathrm{[100~Myr]}-\mathrm{[2~Myr]}$) appears to increase with declining $A$(V). This trend implies that variations in stellar population age have more potential to seriously affect the estimates of the attenuation curve for systems with lower dust content albeit stellar populations younger than 100 Myr. In particular, when $A(\mathrm{V}) < 0.5$, if we estimate the linear background slope and the 2175 Angstrom bump strength for objects with the same attenuation curve but ranging from 2 Myr to 100 Myr, the maximal difference in the results is likely to exceed 0.25 in $c_2$ and 4.5 in $A_\mathrm{bump}$. As also displayed in this figure, random errors become tremendous when $A(\mathrm{V}) < 0.08$, and in this case it is impossible to give any reliable estimate.

The requirement for dust-rich systems tends to be another difficulty in tracing the parameters related to the attenuation curve and hampers our attempt to acquire explicit information about the attenuation curve for galaxies or clusters with low dust content. With this in mind, we solidify the data with IRX $>$ 0.6 (which approximately fulfill $A(\mathrm{V}) > 0.5$) in Figures \ref{hist_c2} and \ref{hist_Abump}, and plot the entire sample by dotted lines for distinction, although for our sample there is no obvious disparity in the results except for NGC~3031, where most of the clusters lie at low attenuation levels.

\subsection{Variations in Conversion Factor between $A$(H$\alpha$)
and $A$(6563)}\label{Sec_disc_Att_Ha_mock}

\begin{figure}[!t]
\centering
\vspace*{-10mm}
\hspace*{-4mm}
\includegraphics[width=1.05\columnwidth]{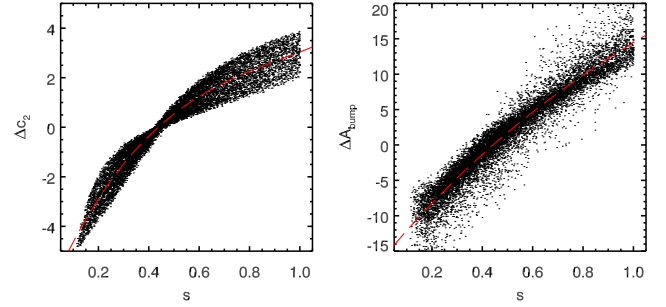}
\vspace*{-12mm}
\caption{$\Delta c_2$ (left panel) and $\Delta A_\mathrm{bump}$
(right panel) as a function of the conversion factor $s$ for the
artificial sample described in Section \ref{Sec_disc_Att_Ha_mock}, where
$\Delta c_2$ and $\Delta A_\mathrm{bump}$ are the differences of the
estimates by using Equations (\ref{Eq_c2}) and (\ref{Eq_Abump}) with
various $s$ from intrinsic $c_2$ and $A_\mathrm{bump}$. The red
dashed lines superimposed on the left and the right panels are the
best-fitting curves for the data, defined as Equations
(\ref{Eq_delc2}) and (\ref{Eq_delAbump}), respectively.}
\label{corr4s_mock}
\end{figure}

In this paper, we employ $A$(6563) in combination with $A$(FUV) and $A$(NUV) to break the degeneracy of the linear background and the 2175 Angstrom bump in UV wavelength bands. The $A$(6563) for observational data is translated from $A(\mathrm{H}\alpha)$ by adopting the conversion $A(6563) = 0.44A(\mathrm{H}\alpha)$ calibrated in \citet{1997AIPC..408..403C}. However, the conversion factor between $A$(H$\alpha$) and $A$(6563) is considered to depend on the geometric configuration between dust, gas, and stars and to vary in different environments. The assumption of the fixed value 0.44 is an empirical simplification and likely to introduce an extra uncertainty into the results. In this subsection, we make an examination of this potential influence on the results.

By denoting the conversion factor as $s$, the conversion becomes
\begin{equation}
A(6563) = s A(\mathrm{H}\alpha). \label{Eq_conversion}
\end{equation}
According to this definition, the assumption of the fixed value 0.44
for the factor $s$ yields the transform of $0.44A(\mathrm{H}\alpha) = 0.44(A(6563)/s)$, where $0.44A(\mathrm{H}\alpha)$ is the derived H$\alpha$-band continuum attenuation and $A$(6563) is the intrinsic H$\alpha$-band continuum attenuation

The difference between 0.44$A$(H$\alpha$) and $A$(6563) as a
function of 0.44$A$(H$\alpha$) is illustrated in Figure \ref{check_Att_Ha_mock} for the artificial sample described in Section \ref{Sec_mock}, where $s$ varies in the range of 0.1$-$1.0. It is obvious to see that in the case of variable $s$, there is likely to be considerable deviation in the estimates of H$\alpha$-band continuum attenuation by assuming $s=0.44$. For instance, if we obtain $0.44A(\mathrm{H}\alpha)=1.0$ mag, a variation of $\pm 0.1$ in $s$ will deviate the result by about $\pm 0.2$ mag, and a variation of $\pm 0.2$ in $s$ will increase the deviation to about $\pm 0.5$ mag.

Figure \ref{corr4s_mock} shows the deviation in $c_2$ and $A_\mathrm{bump}$ as a function of $s$, respectively, for the artificial sample. We also formulate the influences of the variable $s$ on $c_2$ and $A_\mathrm{bump}$ via polynomial fits. The best-fitting curves are superimposed on Figure \ref{corr4s_mock}, and the equations are presented as follows:
\begin{equation}
\Delta c_2 = 7.42 s^3 - 20.67 s^2 + 23.03 s - 6.74, \label{Eq_delc2}
\end{equation}
and
\begin{equation}
\Delta A_\mathrm{bump} = 3.17 s^3 - 15.34 s^2 + 42.86 s - 16.23.
\label{Eq_delAbump}
\end{equation}
Apparently, if the variations in the factor $s$ exceed a certain range, it will have a remarkable impact on the derivation of the linear background slope and the 2175 Angstrom bump area from $A$(H$\alpha$). For instance, a variation of $\pm 0.1$ in $s$ around 0.44 introduces uncertainties of about $-$1.0 to 0.8 in $c_2$ and about $-$3.3 to 2.9 in $A_\mathrm{bump}$, and a variation of $\pm 0.2$ in $s$ around 0.44
increases the uncertainties to about $-$2.3 to 1.5 in $c_2$ and about $-$6.8 to 5.7 in $A_\mathrm{bump}$.

Let us take NGC~3031 and NGC~7331 as an example to describe the influences in two extreme cases. In the previous sections, with $s=0.44$, we have suggested the presence of a prominent bump for NGC~3031 and a bumpless feature for the central ring in NGC~7331. However, for NGC~3031, if the intrinsic value of the factor is $s=0.64$, then the result from the assumption of $s=0.44$ would overestimate $A_\mathrm{bump}$ by about 5.7, and in this case, NGC~3031 appears with a bumpless feature and a very shallow slope. For the NGC~7331 central ring, if the intrinsic value of the factor is $s=0.24$, then the result from the assumption of $s=0.44$ would underestimate $A_\mathrm{bump}$ by about 6.8, and in this case, the ring area has a strong bump in the attenuation curve with a much steeper slope.

\begin{figure}[!ht]
\centering
\includegraphics[width=\columnwidth]{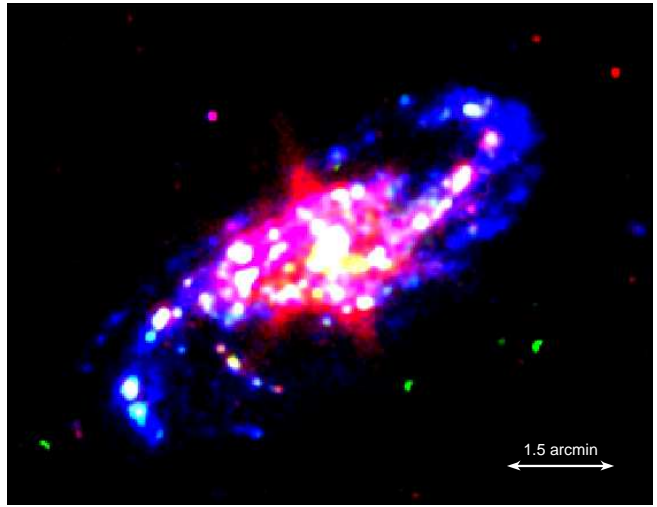}
\caption{Three-color composite image of NGC~4536. The FUV
(blue), continuum-subtracted H$\alpha$ (green) and 24
$\mu$m (red) images are retrieved from \emph{GALEX},
a ground-based telescope, and \emph{Spitzer} observations
respectively. The scale in the bottom right corner indicates the
1$\arcmin$.5 length. North is up and east is to the left.}
\label{RGB_NGC4536}
\end{figure}

A strict calibration of the conversion factor $s$ requires separate estimates of $A$(H$\alpha$) and $A$(6563) from observations. At present, alternative values have been suggested in several studies \citep{2005MNRAS.361..437B, 2009A&A...495..759A}, and as a specific example, with a combination of photometrical SED fitting and spectral emission-line diagnostics, \citet{2013ApJ...769..127L} find the best-fitting value of $s \sim 0.32$ for star-forming regions inside a nearby spiral galaxy M101. The geometric configuration between dust, gas, and stars is suspected to dominate this conversion factor. According to the age-selective attenuation, younger stellar populations tend to encounter heavier attenuation \citep[i.e., $A$(6563) varies with stellar population age.][]{2000ApJ...542..710G, 2007MNRAS.375..640P}, and in this case if we assume a constant geometry between dust and gas in different environments (i.e., $A$(H$\alpha$) is constant), then the value of $s$ will decrease as stellar populations evolve. This scenario can be roughly applied to galaxies as a whole. For subregions in galaxies, the actual geometries seem to be more complicated, and the influences on local geometries between dust, gas, and stars include particle interaction and dynamics, radiative fields, stellar winds and shocks, etc. Separate probes into gas and stars are necessary to recalibrate this conversion.

\subsection{H$\alpha$ Attenuation for NGC~4536}\label{Sec_disc_NGC4536}

As we have mentioned in Section \ref{Sec_ExtCurv}, a majority of the
clusters inside NGC~4536 except the galactic core, appear with a larger amount of H$\alpha$-band continuum attenuation than the expectation from the
MW curve, hinting at shallower slopes than the MW type for these regions (shown in the top left panel of Figure \ref{IRHaUV_Att}). However, this phenomenon conflicts with the $A$(FUV) versus $A(\mathrm{FUV})-A(\mathrm{NUV})$ diagram for this galaxy (the top left panel of Figure \ref{comp_Att}) that requires the MW-type or steeper slopes, otherwise it will have to yield a negative bump (i.e., a trough) in the attenuation curve. The discussion in the above subsection implies that if the conversion factor $s$ differs from the assumed value 0.44, the observed relation between $A$(FUV) and 0.44$A$(H$\alpha$) would be influenced. In consideration of this effect, we find that, when the factor $s$ is adjusted to $\sim$0.25$-$0.3, most of the data points from NGC~4536 are able to fit the MW slope. As a result, the variations in $s$ provide a tentative explanation of the discrepancy between the diagrams in Figures \ref{comp_Att} and \ref{IRHaUV_Att} for NGC~4536.

Notwithstanding, there is another possible reason for this $A$(H$\alpha$) conundrum, which rests in the mismatch between 24 $\mu$m and H$\alpha$ emission from NGC~4536. In this work, $A$(H$\alpha$) is derived from the ratio of 24 $\mu$m to H$\alpha$ luminosities by using Equation (\ref{Eq_Att_Ha}). As we have introduced in Section \ref{Sec_ExtCurv}, the application of Equation (\ref{Eq_Att_Ha}) avails on the basis of the correlation between 24 $\mu$m and intrinsic H$\alpha$ luminosities for star-forming regions (i.e., a constant value for 24 $\mu$m-to-H$\alpha$ in unattenuated environments). However, the measured 24 $\mu$m luminosities for the vast majority of the clusters inside NGC~4536 are suspected to exceed the local star-forming properties. Figure \ref{RGB_NGC4536} shows a composite image of FUV, continuum-subtracted H$\alpha$, and 24 $\mu$m emission for NGC~4536. In this image, the FUV pattern exhibits a clear spiral structure, and the H$\alpha$ knots reside along the spiral arms. The
most conspicuous appearance of this galaxy in this figure is a giant 24 $\mu$m-emitting clump covering from the center to over half of the galactic radius. The intensive central star-forming activities performed in NGC~4536 \citep{2005ApJ...630..837J} are considered as the energy source to heat interstellar dust on this large scale and contribute the prominent IR emission. In contrast to this 24 $\mu$m property, there is no compatible H$\alpha$ source inside this galaxy, and the central star formation is only able to ionize limited regions of gas. As a consequence, most of local 24 $\mu$m sources inside NGC~4536 are likely to be contaminated from the central emission, which potentially induces an overestimate of 24 $\mu$m luminosities in photometry for the local star-forming properties linked to the H$\alpha$ emission (i.e., a larger value for the ratio of 24 $\mu$m to intrinsic H$\alpha$ luminosities). In such a case, Equation (\ref{Eq_Att_Ha}) with the assumption of $a = 0.031$ is suspected to overestimate the H$\alpha$ attenuation, and as a consequence, a majority of the data points from NGC~4536 present higher $A$(H$\alpha$) at fixed $A$(FUV) than any model expectation in Figure \ref{IRHaUV_Att}, resulting in shallower slopes in the attenuation curve. Nevertheless, the 24 $\mu$m excess does not take place in the core of this galaxy where the 24 $\mu$m and the H$\alpha$ luminosities are related to the same energy source, and the location of the central clusters in Figure \ref{IRHaUV_Att} is well fitted by the MW-type slope.

\section{SUMMARY AND OUTLOOK}\label{Sec_sum}

The main purpose of this work is to investigate the features of the attenuation law on both a theoretical and observational basis and thus to offer a complementary interpretation of the dispersion in the IRX-UV relation as a continuation of Paper I, which has discovered the influences of stellar population age.

In this paper, the spectral synthesis modeling predicts the linear background and the 2175 Angstrom bump in the attenuation curve to be two additional parameters in the IRX-UV function; at the same time, the spatially resolved study of the four nearby galaxies, NGC~3031, NGC~4536, NGC~5194, and NGC~7331, provides an observational insight into the features of the attenuation curve.

The linear background with steeper slope, as thoroughly examined in our work, is responsible for redder UV color; the 2175 Angstrom bump, however, tends to diminish the reddening or even have a blueing impact on UV color if it is strong enough. By combining dust attenuations at the three different bands, FUV, NUV, and H$\alpha$, the linear background slope and the 2175 Angstrom bump for the four galaxies are estimated. In the results, NGC~3031 appears with a bump feature in the attenuation curve, but the lack of high attenuation data for this galaxy hinders more accurate constraints. NGC~4536 is well characterized by a bumpless attenuation curve with a MW-type slope similar to the C00 curve in shape. NGC~5194 also presents a bumpless feature in the attenuation curve but with the steeper slope between the MW and SMC types. NGC~7331 exhibits different features in the attenuation curve between the disk and the ring areas, where the disk regions are suggested to have the MW-type bump, and the ring regions manifest an evident absence of a bump in the attenuation curve. The variations in the attenuation curve interpret the deviation in the IRX-UV relation in an additional way and complement the scenarios assuming a constant attenuation curve addressed in Paper I. Our results for NGC~3031 and NGC~5194 are in agreement with earlier studies: \citet{2011AJ....141..205H} have revealed the presence of a MW-type bump in the best-fitting attenuation curves for subregions inside NGC~3031, and \citet{2005ApJ...633..871C} have suggested a quite trivial property of the 2175 Angstrom bump in the attenuation curve for NGC~5194 in contrast to that for the MW.

People have been investigating the deviation in the correlation between the IR-to-UV ratio and the UV color (or UV spectral slope) for a long period of time, yet the features of the attenuation law have not been evidently discovered in observations. In this paper, several possible obstacles that prevent successful discoveries are discussed. The degeneracy of the linear background and the 2175 Angstrom bump in UV wavelength bands is suspected to be one significant cause and makes it impossible to distinguish between the features of the linear background and the 2175 Angstrom bump without suitable observations at other wavelength bands.

Before making efforts to probe and distinguish the parameters in the attenuation curve, people are usually perplexed with variations in intrinsic UV color for different galaxies in more general cases. The aging of stellar populations is a primary cause of the variation in intrinsic UV colors of galaxies and therefore dominates the scatter in the IRX-UV plane, as suggested in a number of papers \citep[e.g.,][]{2004MNRAS.349..769K, 2007ApJ...655..863D, 2009ApJ...703..517D, 2012A&A...539A.145B, 2013ApJ...773..174G} and especially characterized in Paper I. Because of the effects of stellar population age on intrinsic UV color, variations in the attenuation law become a subordinate origin of the deviation in the IRX-UV relation, particularly for composite stellar populations; moreover, even if a sample consists of stellar populations young enough with similar intrinsic UV colors, the features of the attenuation law are expected to be observable only for the dust-rich objects contained therein, since low dust content is still likely to leave the features masked by other effects.

A comparison of attenuations at different wavelength bands is a feasible approach to constraints on the attenuation curve. However, uncertainties in estimates of the attenuations have considerable potential to become an extra obstacle to probing the attenuation law. As a specific example, in this paper, during the estimation of H$\alpha$-band continuum attenuation
($A$(6563)) from H$\alpha$ emission-line attenuation ($A$(H$\alpha$)), we assume the conversion factor of 0.44 as calibrated in \citet{1997AIPC..408..403C}. If this factor in reality happens to be different from 0.44, the resulting A(6563) and the consequent diagnosis of the attenuation curve can be seriously affected, depending on the deviation between the assumed and real values as we have examined and discussed comprehensively in this paper.

Although the attenuation curve is suspected to vary in different galaxies, the connection between the attenuation curve and galaxy type has not been found with robust evidence. In our work, except for NGC~7331 where the central ring area presents a distinct feature from the disk, there is no evident correlation of the parameters in the attenuation curve with spatial locations discovered in other galaxies. At present, physical origins of the linear background and the 2175 Angstrom bump in the attenuation curve remain controversial, and most conclusions about this topic are drawn from experimental products in laboratory or theoretical work by modeling. The linear background appears to correlate with the size distribution of overall dust grains, and in particular small grains with sizes of $\sim 0.01~\mu \mathrm{m}$ are likely to contribute a remarkable rise in the slope \citep{1990A&A...237..215D, 2001ApJ...554..778L,
2004ApJS..152..211Z, 2001ApJ...548..296W, 2003ApJ...588..871C}. The
2175 Angstrom bump is possibly produced by carbonaceous particles, since the $\pi \rightarrow \pi^*$ transition of electrons in carbonaceous molecules triggers the absorption of photons around 2175 Angstrom. Candidates of the bump carrier include amorphous carbon \citep{2011A&A...528A..56G}, Bucky Onion \citep{2005PhRvL..94j5501R}, graphite \citep{2009MNRAS.394.2175P}, and polycyclic aromatic hydrocarbon \citep{1992ApJ...393L..79J, 2004A&A...426..105M, 2006ApJ...639L..59D, 2011ApJ...742....2S}.

In addition to the physical origins, the geometric configuration between
dust and stars is also able to effectively influence the shape of the
attenuation curve. It is widely suggested that dust-star geometries are variable in different radiative field environments, which changes both the linear background and the 2175 Angstrom bump in the attenuation curve \citep[e.g.,][]{2000ApJ...528..799W, 2000ApJ...533..236G, 2001PASP..113.1449C, 2006MNRAS.370..380I, 2007MNRAS.375..640P}. The geometrical effects tend to prevail in measurements of not only galaxies as a whole, but also subregions on several hundreds or tens of parsec scale inside galaxies. In following studies of this series, we plan to investigate the physical and geometrical reasons for the variations in the attenuation law and to attempt to disclose the nature of different features in attenuation curve.

\acknowledgments

We are grateful to the anonymous referee for the careful review and instructive comments that have substantially improved the paper. We thank Robert C. Kennicutt Jr. for the initial discussion with us that helped to draw the general profile of this paper. This work is supported by the National Basic Research Program of China (973 program 2013CB834900), the National Natural Science Foundation of China (NSFC, Nos. 11225315 and 11320101002), the Specialized Research Fund for the Doctoral Program of Higher Education (SRFDP, No. 20123402110037), and the Strategic Priority Research Program "The Emergence of Cosmological Structures" of the Chinese Academy of Sciences (No. XDB09000000). Ye-Wei Mao acknowledges the support of the China Scholarship Council. This research has made use of the NASA/IPAC Extragalactic Database (NED), which is operated by the Jet Propulsion Laboratory, California Institute of Technology, under contract with the National Aeronautics and Space Administration. This research has also made use of the NASA¡¯s Astrophysics Data System.

\vspace*{10mm}

\appendix{\hfill A \hfill}

\section{A TEST OF THE FM PARAMETERIZATION IN THE CASE OF VARIOUS DUST-STAR GEOMETRIES}\label{Sec_App1}

In this paper, we make use of the FM parameterization (Equation (\ref{Eq_FM})) to reproduce attenuation curves considered to have been affected by dust-star geometries. The prescription of Equation (\ref{Eq_FM}) is made from the investigation of individual stars where the geometric configuration is simple and can be assumed as dust screens in foreground of stars, whereas in most cases dust and stars are actually mixed together in various forms and the geometric configuration becomes complex. Effective extinction in complex geometries (termed as \emph{attenuation}) is supposed to differ from the extinction imposed by the foreground screen even for the same dust content and the identical grain types, due to the return of scattered light to the sightline of the observer occurring in complex geometries. In this situation, it is necessary to inspect whether the FM parameterization is valid for various dust-star geometries. This appendix presents a simple test to perform this inspection.

We adopt the MW extinction curve as the receptor of geometric effects because of the conspicuous appearance of the three components (i.e., the linear background, the 2175 Angstrom bump, and the FUV rise) in the MW curve; six types of complex dust-star geometries including three global environments (\emph{shell}, \emph{dusty}, and \emph{cloudy}) plus two local distributions (\emph{homogeneous} and \emph{clumpy}), together with the simple configuration of a foreground homogeneous dust screen, are employed in the test. Attenuation curves affected by these geometries are reproduced from the DIRTY radiative transfer model \citep{2001ApJ...551..269G, 2001ApJ...551..277M} and obtained from \citet{2000ApJ...528..799W}. Definitions of these geometric configurations, i.e., \emph{shell}, \emph{dusty}, \emph{cloudy}, \emph{homogeneous}, and \emph{clumpy}, are addressed in \citet{2000ApJ...528..799W}.

\begin{figure}[!h]
\centering
\includegraphics[width=0.7\columnwidth]{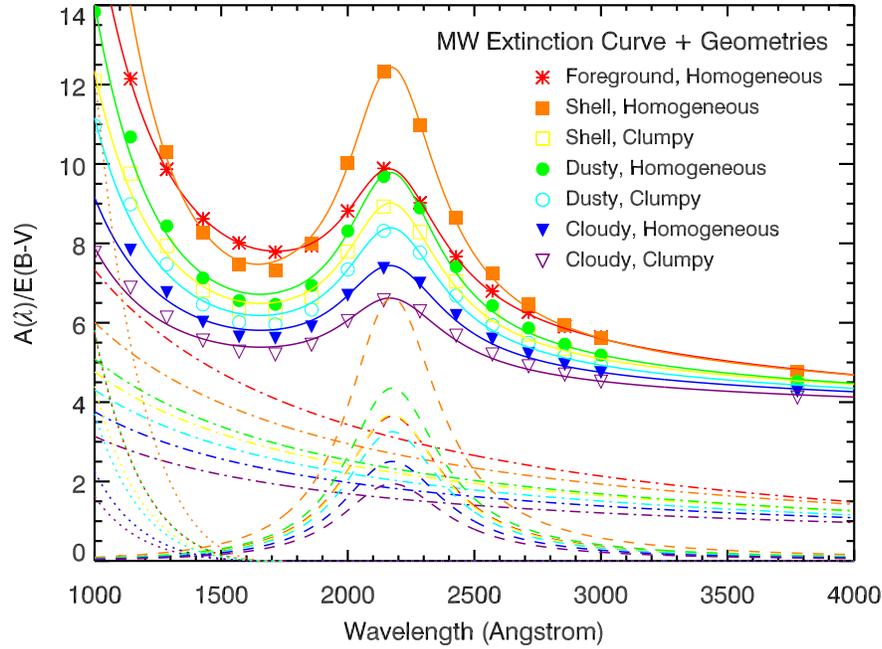}
\caption{Attenuation curves for the MW extinction mode with different dust-star geometries taken into account. The data points are reproduced with seven dust-star geometries as described by the legend in this figure. The solid lines are the best-fitting curves for the data points; three components of the best-fitting curves are also superimposed, including the 2175 Angstrom bump (dashed lines), the FUV rise (dotted lines), and the linear background (dot-dashed lines). All the lines in this figure are color-coded by the seven dust-star geometries corresponding to the data points. The best-fitting coefficients of the attenuation curves are presented in Table \ref{Tab_CurvGeo}.}
\vspace*{10mm}
\label{ExtCurv_mw_Geo}
\end{figure}

We fit the obtained attenuation curves with Equation (\ref{Eq_FM}). Table \ref{Tab_CurvGeo} presents the coefficients of the FM parameterization for these curves, and Figure \ref{ExtCurv_mw_Geo} shows the modeled data and the best-fitting curves. The data points in Figure \ref{ExtCurv_mw_Geo} are the products of the DIRTY model at $\tau_\mathrm{V} = 1.0$; the solid lines are the best-fitting curves for the data points; and the three components of the best-fitting curves are also superimposed, respectively, as the dashed, dotted, and dot-dashed lines.

This figure clearly demonstrates that the combination of the three components prescribed by the FM parameterization is able to successfully reproduce attenuation curves in the case of complex dust-star geometries; all the differences between the attenuation curves with various geometries appear to be the changes in the slope of the linear background, in the strength of the 2175 Angstrom bump, and in the curvature of the FUV rise. This test confirms the validity of the FM parameterization adopted in this paper taking geometric effects into account. For more detailed analyses of the geometric effects, readers are referred to \citet{2000ApJ...528..799W, 2001PASP..113.1449C}.

\vspace*{10mm}

\appendix{\hfill B \hfill}

\section{DISCUSSION WITH THE $R_\mathrm{V}$-DEPENDENT LAW}\label{Sec_App2}

In addition to the FM parameterization we have employed in this paper, the CCM format of the extinction law provides an alternative description of the extinction curve parameterized only with the total-to-selective extinction $R_\mathrm{V}$, which fits the observations from different sightlines for the Milky Way but lacks the support of substantial evidence from extragalactic sources \citep{1989ApJ...345..245C, 1999ApJ...515..128M,
2000ApJS..129..147C, 2003ApJ...594..279G}. In this appendix, we make an additional inspection of this $R_\mathrm{V}$-dependent extinction curve to find out whether or not the galaxies in our sample have this $R_\mathrm{V}$-dependent property in the attenuation curve. Likewise, the CCM law is employed to describe the attenuation features in this work, although it was established by the study of extinction.

\begin{deluxetable}{lccccc}[!h]
\tablecaption{Coefficients of the Attenuation Curves with Different Dust-Star Geometries Taken into Account in Figure \ref{ExtCurv_mw_Geo}} \tablewidth{0pc} \tablehead{ \colhead{Geometry} &
\colhead{$c_1$} & \colhead{$c_2$} & \colhead{$c_3$} & \colhead{$c_4$} & \colhead{$\gamma$} } \startdata
Foreground,~Homogeneous & $-$0.461 & 0.779 & 3.666 & 0.446 & 1.006 \\
Shell,~Homogeneous & $-$0.104 & 0.611 & 5.687 & 0.954 & 0.858 \\
Shell,~Clumpy & 0.079 & 0.468 & 3.536 & 0.340 & 0.961 \\
Dusty,~Homogeneous & $-$0.013 & 0.510 & 4.030 & 0.449 & 0.926 \\
Dusty,~Homogeneous & 0.108 & 0.420 & 3.260 & 0.286 & 1.002 \\
Cloudy,~Homogeneous & 0.197 & 0.356 & 2.722 & 0.174 & 1.087 \\
Cloudy,~Clumpy & 0.245 & 0.289 & 2.209 & 0.125 & 1.137
\enddata
\label{Tab_CurvGeo}
\end{deluxetable}

\begin{figure}[!h]
\centering
\vspace*{-21mm}
\includegraphics[width=0.8\columnwidth]{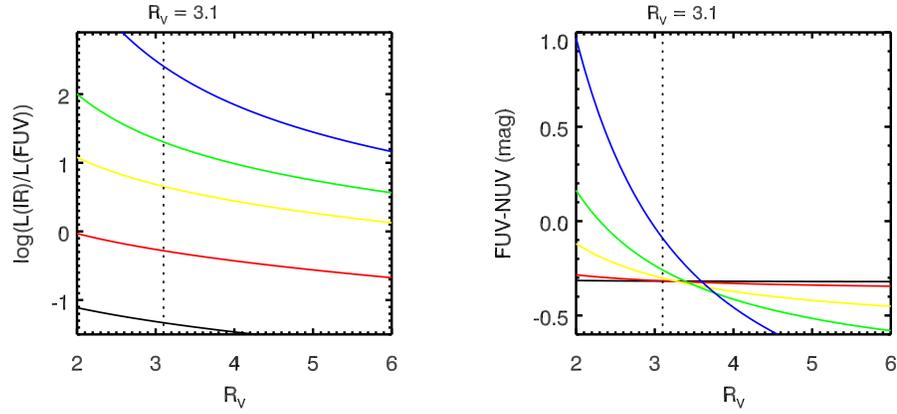}
\vspace*{-23mm}
\caption{IRX (left) and $\mathrm{FUV}-\mathrm{NUV}$ (right) vs.
$R_\mathrm{V}$ reproduced by the CCM extinction law. The solid lines
show the correlations modeled by spectral synthesis at fixed amounts
of dust extinction: $A$(V) = 0.01 (black), 0.1 (red), 0.5
(yellow), 1.0 (green), and 2.0 (blue); the dotted line marks the
position of $R_\mathrm{V}$ in each panel.} \label{Rv_color_vs}
\end{figure}

\begin{figure}[!h]
\centering
\includegraphics[width=0.7\columnwidth]{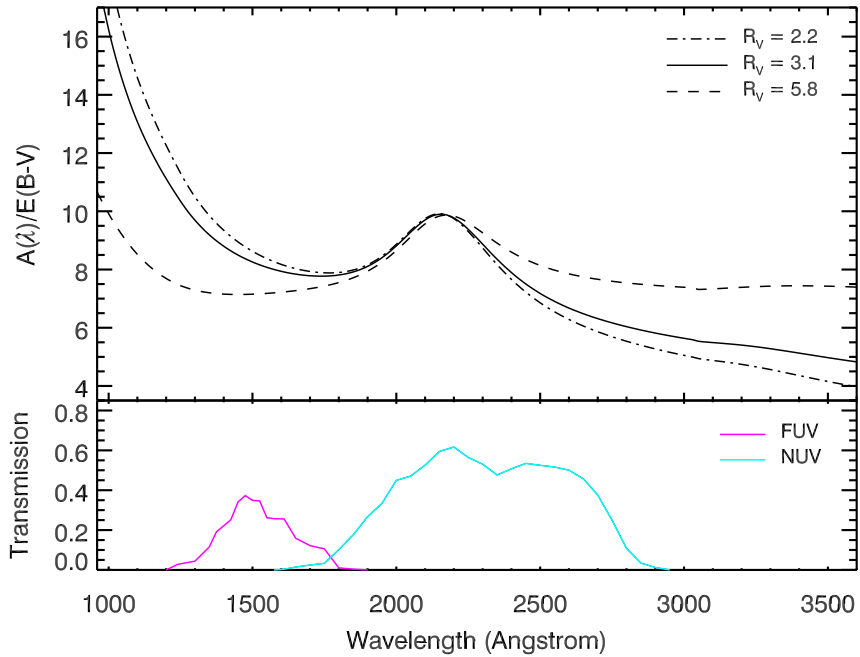}
\caption{Top: examples of the CCM $R_\mathrm{V}$-dependent
extinction curves at the UV wavelength range. The solid line shows
the curve in the case of $R_\mathrm{V}=3.1$; the dashed line and the
dot-dashed line show the curves in the two cases of
$R_\mathrm{V}=5.8$ and $R_\mathrm{V}=2.2$ respectively. Bottom:
filter transmission curves of the \emph{GALEX} FUV (magenta) and NUV
(cyan) bands.} \label{Rv_ExtCurve}
\end{figure}

Figure \ref{Rv_color_vs} shows the variations in IRX and $\mathrm{FUV}-\mathrm{NUV}$ with the parameter $R_\mathrm{V}$ for five constant amounts of $A$(V) (0.01, 0.1, 0.5, 1.0, and 2.0 mag) charged in accordance with the CCM law. Both IRX and $\mathrm{FUV}-\mathrm{NUV}$ decrease with increasing $R_\mathrm{V}$, and the declining trends present steeper gradients at higher attenuation. Particularly, $\mathrm{FUV}-\mathrm{NUV}$ tends to be bluer than intrinsic values when $R_\mathrm{V} \gtrsim 3.5$. In actual observations, $R_\mathrm{V}$ varies in the approximate range of 2.2$-$5.8 with a mean value of $\sim 3.1$ \citep{1999PASP..111...63F}. In Figure \ref{Rv_ExtCurve}, we exhibit the CCM curves with $R_\mathrm{V}$ = 2.2, 3.1, and 5.8 in the UV wavelength portion to show the possible variation trend in the CCM curve within the lower and upper limits. When $R_\mathrm{V}=3.1$, the CCM curve is identical in shape to the MW mean curve of the FM parameterization. With increasing $R_\mathrm{V}$, the slope becomes shallower rapidly, and the 2175 Angstrom bump appears to have a slight reduction in strength. These disparities induce more decreases in FUV attenuation than that in NUV attenuation at larger $R_\mathrm{V}$, which therefore explains the trends of IRX and $\mathrm{FUV}-\mathrm{NUV}$ as a function of $R_\mathrm{V}$ in Figure \ref{Rv_color_vs}.

\begin{figure}[!h]
\centering
\includegraphics[width=0.7\columnwidth]{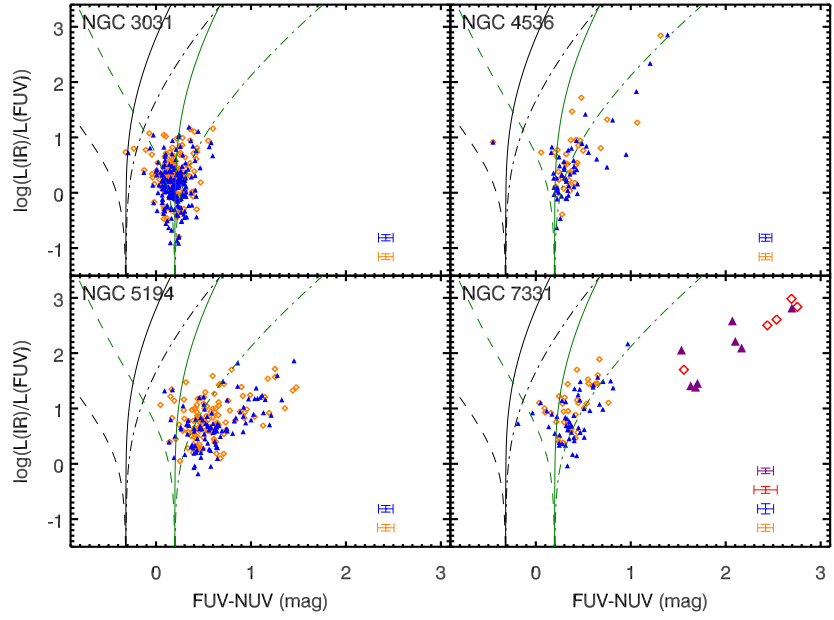}
\caption{Same diagrams for the UV and the IR clusters inside the
galaxies in our sample as plotted in Figure \ref{IRXUV_gal}, but the
lines superimposed on each diagram are reproduced by the CCM
extinction curve in the cases of $R_\mathrm{V}=5.8$ (dashed line),
$R_\mathrm{V}=3.1$ (solid line), and $R_\mathrm{V}=2.2$ (dot-dashed
line) and are color-coded by stellar population age: 2 Myr (black) and
100 Myr (dark green).} \label{Rv_IRXUV_gal}
\end{figure}

\begin{figure}[!h]
\centering
\includegraphics[width=0.7\columnwidth]{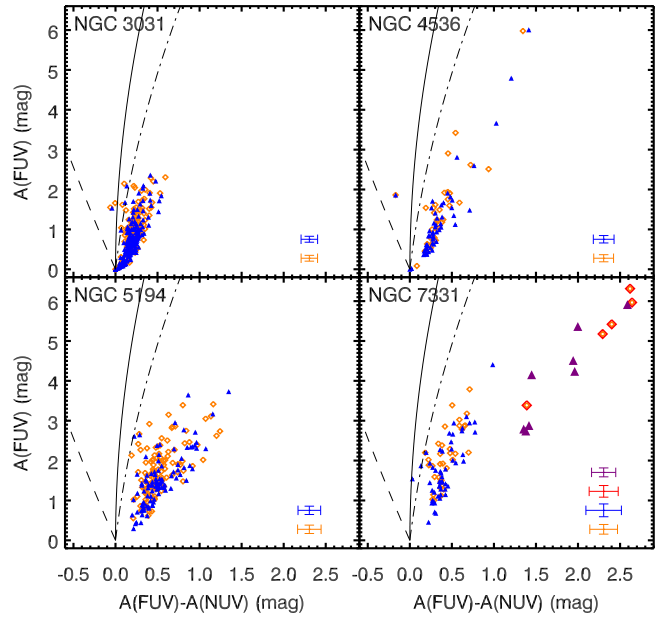}
\caption{Same diagrams for the UV and the IR clusters inside the
galaxies in our sample as plotted in Figure \ref{comp_Att}, but the
lines superimposed on each diagram are reproduced by the CCM
extinction curve in the cases of $R_\mathrm{V}=5.8$ (dashed line),
$R_\mathrm{V}=3.1$ (solid line), and $R_\mathrm{V}=2.2$ (dot-dashed
line).} \label{Rv_comp_Att}
\end{figure}

However, from Figure \ref{Rv_IRXUV_gal} we can see that the CCM law is inappropriate for characterizing the IRX-UV distributions for the galaxies. And in Figure \ref{Rv_comp_Att}, the CCM law within the range of $2.2 < R_\mathrm{V} < 5.8$ fails to interpret the $A$(FUV) versus $A(\mathrm{FUV})-A(\mathrm{NUV})$ relations for the galaxies in our sample, and the data points present larger difference between $A$(FUV) and $A$(NUV) at fixed $A$(FUV) than the model tracks. In the CCM scenario, fitting the
data requires $R_\mathrm{V} < 2.2$ which would break through the observed lower limit. This discrepancy brings forward a caveat that people should be cautious in applying the CCM law to extragalactic environments. The CCM law formulates the three components in the extinction curve as a function of $R_\mathrm{V}$ for our Galaxy, but more studies are required to explore the correlation between the parameters in the attenuation curve and $R_\mathrm{V}$ in general cases.

\end{document}